\documentclass[10pt, conference, letterpaper]{IEEEtran}
\IEEEoverridecommandlockouts

\usepackage{amsthm}
\usepackage{amsmath}
\usepackage{rotating}
\usepackage{hyperref}
\usepackage{tikz}
\usepackage{amssymb}

\usepackage{algpseudocode}
\usepackage{booktabs}

\usepackage{subfigure}
\usepackage{graphicx}
\usepackage{multirow}
\usepackage{color}
\usepackage{xspace}
\usepackage{cite}
\usepackage[algo2e]{algorithm2e}

\usepackage{epsfig}
\newcommand{\BULLET}{\vspace{+.05in} \noindent $\bullet$ \hspace{+.00in}}

\newcommand{\mysubsection}[1]{\vspace{-.05in}\subsection{#1}\vspace{-.01in}}

\newcommand{\smsinglefig}[3]{
\begin{figure}
\centerline{
    \setlength{\epsfysize}{0.18\textwidth}
 \epsffile{\Figdir#1}
} \caption{#2} \label{fig:#3}
\end{figure}
}

\newcommand{\tydubfigsingle}[6]{
\begin{figure}
\centerline{
    \begin{minipage}{0.28\textwidth}
      \begin{center}
        \leavevmode
        \setlength{\epsfxsize}{0.85\textwidth}
        \setlength{\epsfysize}{0.72\textwidth}
        \epsffile{\Figdir#1}
       \newline{\small (a) #2}
      \end{center}
    \end{minipage}
    \hspace*{-1cm}
    \begin{minipage}{0.28\textwidth}
      \begin{center}
        \leavevmode
        \setlength{\epsfxsize}{0.85\textwidth}
        \setlength{\epsfysize}{0.72\textwidth}
        \epsffile{\Figdir#3}
       \newline{\small (b) #4}
      \end{center}
    \end{minipage}
    \vspace{-0.2ex}
} \caption{#5}\label{fig:#6}
\end{figure}
}

\newcommand{\triplefig}[8]{
\begin{figure*}[t]
\centerline{
    \begin{minipage}{0.32\textwidth}
      \begin{center}
        \leavevmode
        \setlength{\epsfxsize}{0.85\textwidth}
        \setlength{\epsfysize}{0.70\textwidth}
        \epsffile{\Figdir#1}\\
       {\small (a) #2}
      \end{center}
    \end{minipage}
    \begin{minipage}{0.32\textwidth}
      \begin{center}
        \leavevmode
        \setlength{\epsfxsize}{0.85\textwidth}
        \setlength{\epsfysize}{0.70\textwidth}
        \epsffile{\Figdir#3}\\
       {\small (b) #4}
      \end{center}
    \end{minipage}
    \begin{minipage}{0.32\textwidth}
      \begin{center}
        \leavevmode
        \setlength{\epsfxsize}{0.85\textwidth}
        \setlength{\epsfysize}{0.70\textwidth}
         \epsffile{\Figdir#5}\\
       {\small (c) #6}
      \end{center}
    \end{minipage}
} \caption{#7}\label{fig:#8}
\end{figure*}
}

\newcommand{\Figdir}{./}

\newcommand{\RN}[1]{%
	\textup{\uppercase\expandafter{\romannumeral#1}}%
}

\newcommand{\bing}[1]{{\color{blue}{[BW: #1]}}}
\newcommand{\vicky}[1]{{\color{brown}{[VM: #1]}}}

\begin{document}

%%
%% The "title" command has an optional parameter,
%% allowing the author to define a "short title" to be used in page headers.
%\title{Continual Learning to Generalize Forwarding in Heterogeneous Mobile Wireless Networks}

%\title{Continual Learning to Generalize Forwarding Strategies Across Diverse Mobile Wireless Networks}

%wb, 6/26/2025
%\title{Generalized Forwarding Strategies for Mobile Wireless Networks: Foundational Models and Fine Tuning}

%\title{Continual Learning to Generalize Forwarding Strategies for Mobile Wireless Networks}

\title{Continual Learning to Generalize Forwarding Strategies for Diverse Mobile Wireless Networks}

%wb, 5/28/2025, the last is the best since we talk about both diverse mobility models, node speed, and transmission range 

%%
%% The "author" command and its associated commands are used to define
%% the authors and their affiliations.
%% Of note is the shared affiliation of the first two authors, and the
%% "authornote" and "authornotemark" commands
%% used to denote shared contribution to the research.

%wb, 7/11/2025
%Author list: Cheonjin Park, Victoria Manfredi, Xiaolan Zhang, Chengyi Liu (tliu02@wesleyan.edu), Alicia P Wolfe, Dongjin Song, Sarah Tasneem (Computer Science) <TasneemS@easternct.edu>; Bing Wang
%

\author{
\IEEEauthorblockN{Cheonjin Park}
\IEEEauthorblockA{University of Connecticut\\
Storrs, CT, USA \\
cheon.park@uconn.edu}
\and
\IEEEauthorblockN{Victoria Manfredi}
\IEEEauthorblockA{Wesleyan University\\
Middletown, CT, USA \\
vumanfredi@wesleyan.edu}
\and
\IEEEauthorblockN{Xiaolan Zhang}
\IEEEauthorblockA{Fordham University\\
Bronx, NY, USA \\
xzhang@fordham.edu}
\and
\IEEEauthorblockN{Chengyi Liu}
\IEEEauthorblockA{Wesleyan University\\
Middletown, CT, USA \\
tliu02@wesleyan.edu} 
\and
\IEEEauthorblockN{Alicia Wolfe}
\IEEEauthorblockA{Wesleyan University\\
Middletown, CT, USA \\
pwolfe@wesleyan.edu}
\and
\IEEEauthorblockN{Dongjin Song}
\IEEEauthorblockA{University of Connecticut\\
Storrs, CT, USA \\
dongjin.song@uconn.edu}
\and
\IEEEauthorblockN{Sarah Tasneem}
\IEEEauthorblockA{Eastern Connecticut State University\\
Willimantic, Connecticut, USA \\
tasneems@easternct.edu}
\and
\IEEEauthorblockN{Bing Wang}
\IEEEauthorblockA{University of Connecticut\\
Storrs, CT, USA \\
bing@uconn.edu}
}

\maketitle

%%
%% The abstract is a short summary of the work to be presented in the
%% article.
\begin{abstract}
Deep reinforcement learning (DRL) has been successfully used to design forwarding strategies for multi-hop mobile wireless networks. While such strategies can be used directly for networks with varied connectivity and dynamic conditions, developing generalizable approaches that are effective on %across 
scenarios significantly different from the training environment  remains largely unexplored. 
%
%can be used in scenarios that differ drastically from training scenarios remain an under-explored area. 
In this paper, we propose a framework to address 
the challenge of generalizability by
% the above generalizability challenge by 
(i) developing a generalizable base model considering diverse mobile network scenarios, and (ii) using the generalizable base model for new scenarios, and when needed, fine-tuning the base model using a small amount of data from the new scenarios. To support this framework, we first design new features to characterize network variation and feature quality,  thereby improving the information used in DRL-based forwarding decisions. We then develop a continual learning (CL) approach able to train DRL models across diverse network scenarios without ``catastrophic forgetting.'' Using extensive evaluation, including real-world scenarios in two cities, we show that our approach is generalizable to unseen mobility scenarios. Compared to a state-of-the-art heuristic forwarding strategy, it leads to up to 78\% reduction in delay,  24\% improvement in delivery rate, and comparable or slightly
higher number of forwards.
\end{abstract}

\section{Introduction}
\label{sec:introduction}

%wb, 6/23/2025, why use fine-tuning in LLM
\iffalse 
The pretrained model was already trained on a dataset that has some similarities with the fine-tuning dataset. The fine-tuning process is thus able to take advantage of knowledge acquired by the initial model during pretraining (for instance, with NLP problems, the pretrained model will have some kind of statistical understanding of the language you are using for your task).
Since the pretrained model was already trained on lots of data, the fine-tuning requires way less data to get decent results.
For the same reason, the amount of time and resources needed to get good results are much lower.
For example, one could leverage a pretrained model trained on the English language and then fine-tune it on an arXiv corpus, resulting in a science/research-based model. The fine-tuning will only require a limited amount of data: the knowledge the pretrained model has acquired is “transferred,” hence the term transfer learning.

Fine-tuning a model therefore has lower time, data, financial, and environmental costs. It is also quicker and easier to iterate over different fine-tuning schemes, as the training is less constraining than a full pretraining.
\fi 

%%wb, 6/26/2025
%Mobile wireless networks are everywhere; including in realworld scenarios; generalized forwarding strategy: foundational models + fine tuning

% Multi-hop 
Mobile wireless networks have been used for many applications, 
including %e.g., 
emergency response, environmental monitoring, and military communications. %deployments. 
Forwarding strategies for such networks
%mobile wireless networks
aim 
% to find network paths to 
% deliver data packets from source to destination. 
to identify good paths over which to 
deliver data 
packets from source to destination. 
% \tommy{Forwarding strategies in wireless networks aim to identify optimal paths for delivering data packets from source to destination.}
\iffalse
Many forwarding strategies have been developed  (see \S\ref{sec:related-work}). 
Some are for connected networks, where despite node movement, end-to-end paths exist between two nodes, and 
%the forwarding strategies use a protocol to find the path and forward the 
these paths typically only exist for short periods of time and hence need to be updated periodically~\cite{}. 
%re-established using ad hoc routing protocols~\cite{}. 
Some are for delay-tolerant networks~\cite{}, where contemporaneous end-to-end paths rarely exist, and packets are forwarded through node mobility. 
\fi
While many forwarding strategies have been developed, they typically
%(see \S\ref{sec:related-work}),
target specific kinds of network connectivity.
In ad hoc networks, for example, despite node movement, end-to-end paths often exist %between node pairs  
but only for short periods of time and so must be periodically updated~\cite{Johnson96dynamicsource,perkins2003rfc3561,clausen2003optimized,Perkins94:DSDV}. 
In comparison, in delay-tolerant networks~\cite{Jain2004:DTN}, contemporaneous end-to-end paths rarely exist and so packets must be forwarded hop-by-hop.

\iffalse
In reality, a mixture of good and poor
connectivity may co-exist in a mobile wireless network due to node movement.
%can be mostly disconnected, while other part may exhibit good connectivity. In addition, 
%Consider a particular node. Its connectivity may change over time.
The connectivity of a node may change dynamically over time.
%, varying between poor and good connectivity due to movement of nodes in the network. 
%leading to good connectivity during some time, while poor connectivity during other time scenarios. 
%In addition, 
At one point of time, it can be difficult for a node to determine whether a %contemporaneous end-to-end 
path exists between itself and a destination  based on local neighborhood information. As a result, it is difficult for a node to switch between different forwarding strategies based on its current connectivity as in~\cite{}.  
\fi

In reality, a mixture of good and bad 
connectivity may co-exist in a mobile wireless network, varying as a function of node movement~\cite{manfredi2011understanding}. Because network connectivity can change over time and space, it can be difficult for a node to ascertain whether a viable path exists to a packet's destination
%determine whether a path exists 
%from a node to a packet's destination
based on %the node's
% between the node %\tommy{source} 
% and a destination based on the its %node's \tommy{its} 
local neighborhood information; this then makes it difficult for the node to determine the right forwarding strategy.
Deep reinforcement learning (DRL), which combines RL \cite{sutton2018reinforcement} with deep neural networks (DNNs) \cite{bengio2009learning}, is one approach to design forwarding strategies able to adapt to the changing network connectivity.
%of a mobile wireless network. 
% With DRL, a node  feeds a set of features describing the  node's state and next hop actions into a DNN, which generates outputs to guide the node's forwarding decisions.
% \tommy{
With DRL, each node inputs a set of features characterizing its current state and possible next-hop actions into a DNN, which then generates outputs to inform forwarding decisions.

%characterized by a network topology that is changing both spatially and temporally. In such networks, devices may meet infrequently which limits the ability of devices both to forward data packets and exchange control state. Consequently, hop-by-hop forwarding must be used as the end-end paths needed for routing rarely exist. Even with hop-by-hop forwarding, however, any control state that devices are able to collect may be out-of-date by the time a forwarding opportunity arises, introducing additional uncertainty into decision-making. \vicky{Diversity of movement ....}

\iffalse
Deep reinforcement learning (DRL), which combines RL \cite{sutton2018reinforcement} with deep neural networks (DNNs) \cite{bengio2009learning}, is one approach to design forwarding strategies %decision-making 
for diverse wireless networks. % despite the connectivity. 
With DRL, a node %simply 
feeds a set of features describing the current node's state and next hop actions
% that are important for making forwarding decisions 
into a DNN, which generates outputs to guide the node's forwarding decisions.
%outputs the forwarding decision.
%for that node. 
\fi

While DRL has been successfully used to design routing and forwarding strategies for a number of different network settings (see \S\ref{sec:related-work}), the strategies are often tested in scenarios that are somewhat similar to the training environment. For instance, although the study in~\cite{manfredi2024learning} develops DRL forwarding strategies across different network scenarios, including different transmission ranges and mobility models, its scope is limited to homogeneous networks.
%
%the study in~\cite{manfredi2024learning} focuses on developing generalizable forwarding strategies across different network scenarios, including different transmission ranges and mobility models. 
% However, the diversity of scenarios in~\cite{manfredi2024learning} is limited with only homogeneous  networks considered. 
%However, the scope of that study is limited to homogeneous network scenarios.
% \tommy{However, the scope of that study is limited to homogeneous networks.}
%For instance, the mobility models include random way point and Gaussian Markov models, both including a group of i.i.d. nodes.
As such, the trained strategies are not suitable for use in  
% drastically different mobility scenarios.
highly diverse mobility scenarios.

%and heterogeneous networks.
%
%(e.g., group-mobility mode) and heterogeneous nodes (e.g., nodes with different average speed, or  belonging to different groups).
%when there are multiple groups). 

An ideal generalizable forwarding strategy should function well in any mobile wireless network, including those it has never encountered in training. One way to achieve this generalization is through  online training, i.e., retraining a model on the fly when encountering a new network scenario (see \S\ref{sec:related-work}). Online training, however, takes time to converge, particularly in sparse networks where nodes have few opportunities to meet each other, and its performance 
%during online training 
tends to be poor before convergence. In addition, it can incur significant computational overhead, 
not suitable for devices with limited resources.
%computation capabilities. %Furthermore, in distributed environment, devices either are limited to local information or require significant 

%wb, 7/26/2025, remove {\em zero-shot} 
In this paper, we are particularly interested in forwarding strategies that are trained {\em offline}, where a model is trained beforehand and then used directly in a new scenario. This is a significant challenge,  considering the potentially infinite number of possible mobile network scenarios. 
%In this paper, we are particularly interested in {\em zero-shot} forwarding strategies, where a model is trained beforehand and then used directly in a new scenario. 
%
%Inspired by the foundation models in natural language processing, 
We propose a framework to address this challenge by first developing a {\em generalizable base model} %offline, 
trained on 
diverse mobile network scenarios. We then
use this base model for {\em forwarding in new scenarios}; when needed, we {\em fine-tune} the base model using only a small amount of data from
a new scenario, and then use it for forwarding
%the fine-tuned model 
in similar scenarios.
%(e.g., fine-tune base model trained using synthetic mobility model for real road networks). 
%deploy apply which is possible since the.
%wb, 7/25/2025, emphasize offline training, which is an important difference from existing studies 
%too detailed; move it to later
%
%While the models are used in a distributed and online manner, both the base model training and fine-tuning are performed offline centrally, leading to much more efficient training than online training.
%
%Both base-model training and model fine-tuning are conducted offline in a centralized manner, and then used online  model leveraging global information, which leads to significantly more efficient training than online training.  
%
%again in an offline manner, and then use the fine-tuned model for similar scenarios.
\iffalse
We propose a framework to address this challenge by (i) developing a generalizable base model trained on %considering 
%a large number of 
diverse mobile network scenarios, and (ii) 
%when encountering a new scenario, 
using the base model to determine forwarding strategies for a new scenario, and when needed, fine-tuning the base model using a small amount of data from
%short trace from 
%corresponding to 
the new scenario.
\fi
%
%and then use the fine-tuned model zero-shot in similar scenarios. 
%
We identify two key components for developing a generalizable base model: 
(i) include a {\em rich set of features} to characterize network variation and feature quality, and (ii) use {\em continual learning (CL)}~\cite{Lange2022:CL,Khetarpal2022:CL,Wang2024:CL} to adapt to diverse network scenarios. 
% We further present real-world use cases in two cities.
\iffalse
(i) include a rich set of features to characterize network variation and feature quality, and (ii) use {\em continual learning (CL)} to adapt to diverse network scenarios. We further present real-world use cases in two cities.
\fi
%for using the base models.
%both inspired by real-world mobility scenarios. 
%One use case does not require fine-tuning, while the other requires fine-tuning.  
Specifically,  we make the following  contributions:

\iffalse 
forwarding decisions are able to easily take into consideration a variety of disparate features derived from the exchanged control state, including features that explicitly model feature quality. Yet, while DRL has been used to design routing and forwarding strategies for a number of different network settings (see \S\ref{sec:related-work}), 
\vicky{using DRL to ... }

\vicky{To generalize a forwarding strategy across diverse networks ... continual learning}
\bing{Mention creating a base model using diverse settings through continual learning. }
\bing{Two real-world traces: Manhattan and SUMO Bogna settings.}

In this work, we .. We first design new features to characterize network variation and feature quality, and thereby improve the information used in DRL forwarding decisions. 
\vicky{Then, we design continual-learning ... }
Using simulation under a wide range of settings, we evaluate the performance of our approach. Specifically,  we make the following  contributions.
\fi 

\BULLET {\em Characterization of diverse mobile wireless networks (\S\ref{sec:features}).} We 
propose %introduce 
three new classes of features to characterize network diversity: %, specifically 
(i) quality-of-information metrics, 
(ii) network memory metrics, and 
(iii) community metrics. 
% The quality metrics are helpful in assessing the quality of the information being used in DRL decision-making. The other two 
Features in these classes %additionally
evaluate information about the network over 
% different periods of 
time, and hence can provide valuable information  even when instantaneous features have become out-of-date. 
We show that these features are important for characterizing heterogeneous %diverse
network conditions, and 
for training a base 
forwarding 
model that is generalizable
% for training an effective base forwarding model
%DRL-based forwarding strategies 
across diverse network scenarios.

\BULLET {\em Generalizable policy trained using CL (\S\ref{sec:drl_forwarding}).} 
We address %resolve 
the  challenge of generalizing DRL-based forwarding strategies to disparate network scenarios through the use of CL.
%We resolve the generalizability challenges of training DRL-based forwarding strategies to disparate network scenarios through the use of CL. 
%
We design a simple and robust  method that learns generalizable forwarding strategies using CL from multiple diverse network scenarios without suffering from ``catastrophic forgetting''.

%We show that the CL-trained %continually learned 
%models are indeed more generalizable, without suffering from ``catastrophic forgetting'' and can achieve good performance even in new mobility scenarios.  
% These models lead to an even greater performance advantage over non-DRL heuristics in large networks, where routing is even more challenging, versus in %than in 
% small networks. 
%In addition, the continually learned models demonstrate a greater performance advantage over non-DRL models in larger-scale networks, where routing is more challenging.

\BULLET {\em Extensive evaluation (\S\ref{sec:evaluation}) and real-world scenarios (\S\ref{sec:real-world}).} Using evaluation in 21 small and 36 large synthetic network scenarios, as well as 18 real-world  
scenarios in two cities that cover a wide range of network density, connectivity and mobility models, we show that  our approach is generalizable. It outperforms a state-of-the-art heuristic  strategy in most scenarios, achieving up to 78\% reduction in delay, 24\% improvement in delivery rate, and incurring comparable or slightly more forwards.

\iffalse 
We evaluate the DRL forwarding strategy learned using CL in 
%a wide range of scenarios, including 
%very sparse and very dense networks, evaluating 
21 small and 36 large synthetic network scenarios, as well as 18 real-world %road  networks 
%scenarios set 
scenarios
in two cities. These scenarios are in a wide range, differing significantly in network density, connectivity and mobility models. 
The evaluation results show that our approach is generalizable. It outperforms a state-of-the-art heuristic  strategy in most scenarios, providing up to \bing{74\%} reduction in delay, \bing{24\%} improvement in delivery rate, and comparable or slightly higher number of forwards.
\fi 

\iffalse 
We evaluate our forwarding strategy over a variety of network scenarios, 
\color{red}
training a DRL agent on a small network of 25 devices in a small area, and testing it on a large network of 100 devices operating in a larger area ...
\color{black}
\vicky{emphasize real-world scenario}
We show that the model is generalizable to 36 large-scale network scenarios with  unseen  mobility patterns (\S\ref{sec:evaluation}, they).
%, with average node degree to 0.1 to 15.8. 
Furthermore, with fine-tuning in one real-world scenario in a single city, the model is generalizable to 18 diverse real-world scenarios,  including those from a different  city (\S\ref{sec:real-world}).
\fi 
\section{Related Work}
\label{sec:related-work}

% \begin{table}[t]
% \centering
% \caption{{Constants used in our feature estimations and simulations. \bing{Maybe we can omit this table; only describe this in the text to save space.} \vicky{I can address describing in text}}}
% \begin{tabular}{lll}
% \toprule
% {\bf Name}      & {\bf Meaning}                & {\bf Value} \\
% \midrule
% % $N$             & Number of devices in network     & 25, 100 \\
% $B$             & Maximum buffer length        & 200 \\
% $T$             & Initial time-To-Live (TTL)   & 300 \\
% $D$             & Maximum node degree          & 10 \\
% $L$             & Maximum Euclidean distance   & 1500 \\
% $k$             & Small normalization constant & 0.001 \\
% $k^\prime$      & Medium normalization constant & 0.01 \\
% $k^\prime\prime$ & Large normalization constant & 0.1 \\
% \bottomrule
% \end{tabular}
% \label{tab:sim-constants}
% \end{table}

%wb, 3/12/2024, di2019carma actually uses RL instead of DRL
%omit some low-quality papers: Mukhutdinov19:Multi-agent, Suarez-Varela19:feature, ye2015multi, you2019toward (conference version of you2020toward, so only includes the journal version)
%vehicular networks: remove schuler2021robust,
%UAV network: remove schuler2021robust
%general mobile network: remove jianmin2020ardeep (4-page paper with limited contribution)

%wb, remove Li2022:survey for space
\noindent{\em RL and DRL for routing.} Q-routing~\cite{Boyan94:Q-routing} is the first scheme that uses RL \cite{sutton2018reinforcement}
for routing. Since then, many more schemes that use RL or DRL 
%for routing 
have been proposed (see surveys~\cite{Al-Rawi2015:review,Mammeri19:RL,Luong2019:survey,Mao2018:comprehensive}). 
%
\iffalse 
Many follow-up studies (see surveys~\cite{}), as in~\cite{Boyan94:Q-routing}, use  look-up tables, and hence are not scalable.
%It is extended by~\cite{Choi96:PQ-routing,Kumar98:Confidence,hu2010adaptive,elwhishi2010arbr,rolla2013reinforcement,di2019carma}. These approaches  use look-up tables, and hence are not scalable. 
%which do not scale and so are unsuitable for complex domains such as mobile wireless networks. 
Recent advances in deep learning motivate DRL-based routing~\cite{Valadarsky17:learning,Xu2018:experience,you2020toward,chen2021multiagent,wowmom2021-relational,almasan2022deep}, which uses a deep neural network (DNN) as a %value 
function approximator to replace look-up tables. 
\fi 
%
Existing DRL-based routing and forwarding strategies for mobile wireless networks typically target specific network types, % \tommy{typically target specific network types}, 
e.g., vehicular networks \cite{li2018hierarchical,lolai2022reinforcement,luo2021intersection},  UAV networks \cite{feng2018multi,sliwa2021parrot,rovira2021fully,qiu2022data},  IoT networks~\cite{sharma2020rlproph}. %community networks~\cite{han2021qmix},%wb, 7/30/2025 
%or other limited settings (e.g., with a small number of flows or  devices)~\cite{johnston2018reinforcement,sharma2020rlproph,kaviani2021robust}. 
In addition, most 
studies focus on online learning (e.g., as in Q-routing~\cite{Boyan94:Q-routing}), which is much less efficient than offline training.
%which can use all of the data collected from all of the nodes.
%
%particularly for
%leads to poor performance before convergence and is not feasible in 
%sparse mobile networks due to the limited opportunities to exchange information. 
The study most closely related to ours is~\cite{manfredi2024learning}, which, like our work, employs offline training.
% The work that is closest to ours is~\cite{manfredi2024learning}, which uses offline training, as does our study. 
% \tommy{The study most closely related to ours is~\cite{manfredi2024learning}, which, like our work, employs offline training.} 
We nevertheless examine a significantly broader range of networks (including real road networks) than does~\cite{manfredi2024learning}, and focus on improving generalizability across diverse networks by proposing novel features and developing CL-based approaches. 

\smallskip
\noindent{\em CL.} Many CL techniques have been developed that allow training of an intelligent system using new tasks without forgetting what was learned in earlier tasks~\cite{Lange2022:CL,Khetarpal2022:CL,Wang2024:CL}. We adopt a replay-based technique to our problem setting.
%use a replay-based technique that can be easily incorporated into our setting. 
While CL has been used in many applications, only a few studies explore its use in computer networks. The study in~\cite{LozanoCuadra2025} uses CL to adapt routing to time-varying conditions (e.g., traffic, link states, and mobility patterns) in Low Earth Orbit satellite networks. It employs CL in an online manner, whereas our study uses CL in an offline manner. Other studies use CL for other contexts, such as resource allocation~\cite{Sun2022:CL}.
%, channel estimation~\cite{}. 

\iffalse 
\noindent{\em Continual learning for routing.} 
\color{blue}
A lot of work on continual learning.
What work has looked at for routing and for graphs?
\color{black}
\fi

\smallskip
\noindent{\em Characterizing heterogeneous mobile networks.}
% \bing{Vicky, can you shorten this part? Maybe just emphasize that we identify new classes of features. Some are existing, but have not been used as features for DRL. We also design/adapt new features that can be easily estimated in mobile wireless networks.} 
In this work we identify new classes of features to characterize heterogeneity in mobile networks.  While some of the features that we propose for these classes have been identified in previous works (see~\cite{ferriere2003agematters,singlecopy-spyro,yates2021age} and \S\ref{sec:features}), they have not previously been used in a DRL context. We additionally design new features for these classes, particularly the short and long dispersion features in \S\ref{sec:memory-features}. For all features, we design methods for easy online estimation in deployed mobile  networks.
\section{Background and High-level Approach}
In this section, we briefly overview DRL, and then present the main ideas in our DRL-based approach to generalize forwarding strategies across diverse mobile wireless networks.

\subsection{DRL Background} 
\label{sec:background}

In RL \cite{sutton2018reinforcement}, an agent interacts with its environment to learn a policy, i.e., the best action for each state of the environment. A  Markov decision process (MDP) is used to model this policy, defined as a 4-tuple ($\mathcal{S}$, $\mathcal{A}$, $\mathcal{P}$, $\mathcal{R}$) where $\mathcal{S}$ is the set of states and $\mathcal{A}(s)$ are the actions available in each state $s \in \mathcal{S}$, while $\mathcal{P}(s,a,s^\prime)$  is a function giving the probability of transitioning from state $s \in \mathcal{S}$ to $s^\prime \in \mathcal{S}$ after taking action $a \in \mathcal{A}$ and $\mathcal{R}(s,a,s^\prime)$ is a function  giving the corresponding reward. In RL, $\mathcal{P}$ and $\mathcal{R}$ are not known, but samples can be obtained via interactions with the environment and used to learn a Q-value  for each $(s, a)$ pair, %where 
$s\in \mathcal{S}$ and $a \in \mathcal{A}(s)$.  This Q-value estimates the expected future reward that the RL agent will receive after taking action $a$ in state $s$;  the action with the highest Q-value is then the  optimal action to take in that state. For large state and action spaces, a DNN can be trained to approximate the mapping from $(s, a)$ to Q-value, as in DRL, where states and actions are described using feature sets. 

\subsection{DRL-based Forwarding Strategies} \label{sec:bk-DRL-forwarding}

%%wb, 7/11/2025, move the details to later on; only talk about packet agents, train policy offline and use it independently, reward, structure of DNN; everything related to features is deferred to later on
In principle, DRL can learn a forwarding strategy that is able to seamlessly adapt forwarding decisions to diverse network conditions by varying the feature values that are input into the associated neural network. This is unlike traditional adaptive forwarding approaches~\cite{danilov2012adaptive, seetharam2015routing,asadpour2016route},
%
%~\cite{danilov2012adaptive, seetharam2015routing, lakkakorpi2010adaptive, delosieres2012batman, raffelsberger2014combined, asadpour2016route}, 
which must first explicitly identify the current network setting and then switch to using the appropriate forwarding strategy for that setting. However, although it is relatively easy to train a DRL forwarding strategy in one network setting and obtain good testing performance in the same setting, training a strategy that performs effectively across diverse settings is more difficult given the time and resource-intensive nature of training.   

\smsinglefig{figs/DRL-packet-agent}{Overview of DRL forwarding strategy of \cite{manfredi2024learning} 
 % \cite{wowmom2021-relational,manfredi2024learning} 
on which we build. 
%\bing{Simplify the figure to remove $f_s()$ and $f_a()$.} 
}{packet-agent-DRL}

A key objective of using DRL in this work is thus to ensure {\em generalizability} of the forwarding strategy, so that the strategy performs well across diverse network settings, including those on which it was not trained. To do so, we build on~\cite{manfredi2024learning} in which DRL agents are defined for {\em packets} rather than devices, and so packets themselves choose their next hops, simplifying learning.  Features are then computed from a packet's viewpoint and learning tracks packet states over time. Importantly, this forwarding strategy is able to be trained offline (so can be used directly
%\vicky{zero-shot} 
in testing scenarios) and the same strategy can be used independently by all packets. 

Fig. \ref{fig:packet-agent-DRL} illustrates the DRL formulation in~\cite{manfredi2024learning}. Each node $v$ in the network repeats the process of exchanging features with neighbors, selecting a packet $p$ to forward, and having $p$ choose a next hop action $u$. 
Packet $p$ separately considers each possible action $u$ available in its current state $v$ by feeding the corresponding state and action features (see more in \S\ref{sec:features-background}) into the DNN that encodes the forwarding strategy.
The DNN then outputs a Q-value measuring the goodness of $p$ moving from $v$ to $u$, and the action with the highest Q-value is selected. 

\subsection{High-level Approach for Generalizability}
Our approach for designing a DRL-based generalizable forwarding strategies has two parts: (i) {\em train a generalizable base model} and (ii) {\em fine-tune the base model when necessary}.
%is twofold: (i) train a generalizable base model over various mobile wireless networks to ensure effective performance in a large number of diverse scenarios, and (ii) use this generalizable base model directly, or when needed, fine-tune it for new scenarios and then use the fine-tuned model zero-shot in those scenarios.
% For scalability, 
%Our approach is similar in spirit to foundation models in natural language processing: 
    
To train a { generalizable base model}, we leverage a combination of {\em novel features}, {\em continual learning (CL)}, and training on {\em diverse} scenarios.
The performance of DRL-based forwarding depends heavily on the input features. We therefore design new features, beyond those in \cite{manfredi2024learning}, to capture node diversity in network settings (see \S\ref{sec:features}). 
Training the model sequentially across multiple diverse scenarios can suffer from ``catastrophic forgetting'', i.e., models trained
on later scenarios do not perform well on earlier scenarios.
%i.e., the model ``forgets'' earlier scenarios when training it on a new scenario. 
We thus design CL-based forwarding strategies to achieve both plasticity and stability (see \S\ref{sec:drl_forwarding}).
Finally, %to train the base model, 
we train the base model on %consider 
%leverage {\em large-scale synthetic datasets} (each with $80$k timesteps), which 
a wide range of settings and large amount of 
% training 
data ($80$k timesteps for each setting), 
%allowing us to judiciously select the mobility models and parameters for mobile wireless networks to 
ensuring %that 
the model is exposed to 
adequate diversity.
% an adequate level of diversity.
%``sees'' sufficient amount of diversity.
%of settings (see Table \ref{tb:training-cases}).
While CL allows the use of many training scenarios, we shall see that, with careful selection, only a few are needed
%three 
to derive a generalizable base model.  

In our evaluation, we find that it is often sufficient to use this base model directly in new scenarios.
%wb, 7/28/2025
%
%(it is generalizable to the 21 small and 36 large network scenarios that we evaluate).
%aka zero-shot learning).
When necessary, we {\em fine-tune} this base model with a small amount of data from a new scenario. % and then use the updated model zero-shot in those scenarios.
For instance, we use 1000 timesteps of data from a real road network in one city to fine-tune the base model, and then use it in 18 road network scenarios from two cities. This fine-tuning is one form of transfer learning~\cite{Pan2010:transfer-learning}: it only uses a small number of samples since it applies the knowledge gained from the diverse datasets contained in the base model.

\subsection{Features for DRL-based Forwarding} \label{sec:features-background}

%The performance of DRL-based forwarding depends heavily on the input features.  
Table \ref{tab:features} lists the %the set of 
features used in \cite{manfredi2024learning} and the new features (in bold) that we introduce and use in this work. %for generalization across diverse networks. 
Nodes exchange features only with their neighbors to limit the control overhead incurred by these exchanges. We briefly describe the features in \cite{manfredi2024learning} below; the new features 
%we introduce 
are deferred to \S\ref{sec:features}.

\iffalse 
$f_{s} (v, p, d) := {f}_{packet}(p) \cup {f}_{direct}(v, d) \cup f_{indirect}(v, d) \cup {f}_{nbrhood}(v, d)$. 
\fi  

 Consider a packet $p$, with destination $d$, 
 % that is 
 currently at device $v$ with neighbors $Nbr(v)$. The {\em state features} for $p$ are defined at multiple levels. Packet-level features capture information about $p$ like $p$'s time-to-live (TTL). Device-level features capture information about $v$ like $v$'s queue length, node degree and node density. Path-level features capture information about the time-varying path from $v$ to  $d$, like the number of packets at $v$ with destination $d$, the estimated and possibly out-of-date Euclidean distance from $v$ to $d$, and the transitive timer for $v$ to $d$ (computed as the time elapsed since  $v$ last met $d$, updated via timer transitivity~\cite{singlecopy-spyro}). Neighborhood-level features are then aggregate values  (minimum, maximum, and mean) computed over the device and path-level features for  $Nbr(v)$. The {\em action features} describe a possible next hop action  $u \in Nbr(v) \cup v$ for packet $p$,  i.e., moving to node $u$ or staying at the current node $v$, and comprise the device-level and path-level features for $u$ along with context features indicating whether $p$ has recently visited $u$ and  whether the action type is to move to a neighbor or stay at the current node.  
% As shown in Fig. \ref{fig:packet-agent-DRL}, packet $p$ separately considers each possible next hop action $u$ for the current state, feeding the state features and corresponding action features to the trained DNN, and selects the next hop that leads to the highest Q-value.

Although the above features 
were shown to be able to characterize a network's 
%current 
conditions and generalize to different mobile network scenarios~\cite{manfredi2024learning}, they also have some critical limitations. 
First, they do not quantify the {\em quality or freshness of feature values}. When some features are of low quality or stale (e.g., not updated for a long time due to sparse network connectivity), the model might be better off ignoring them.
%such features. 
Second, they fail to account for varying  network conditions, such as nodes moving at different speeds or networks with both sparse and dense areas of  connectivity. These limitations motivate  the  new features we introduce 
in \S\ref{sec:features}. %, we address these limitations by introducing new features to characterize diverse network conditions and describe approaches to initialize and normalize feature values to ensure generalizability.

\section{New Features for Diverse Networks} 
\label{sec:features}

% To represent and differentiate diverse mobile wireless networks,  we introduce three classes of temporal features:  {\em quality of information} metrics,  {\em network memory} metrics, and {\em community} metrics. For each class, we propose  specific features (i.e., the features in bold in Table~\ref{tab:features}); exploring other possible features for these classes is left as future work. In the following, we describe each of these features, and discuss their benefits. At the end of this section, we describe initial feature values and feature normalization.

To characterize diverse mobile wireless networks,  we introduce three classes of features:  {\em quality of information} metrics, {\em network memory} metrics, and {\em community} metrics. For each class, we propose  specific features (shown in bold in Table~\ref{tab:features}); exploring other possible features is left as future work. In the following, we describe these features and discuss their benefits. At the end of this section, we describe how we initialize and normalize feature values to ensure generalizability.

%wb, 7/15/2025, updated table
\begin{table*}[t]
\centering
\caption{
\small{Features for packet $p$ at time $t$ with destination $d$  located at device $v$ with possible next hops $u \in Nbr(v) \cup v$, where $x$ specifies the value of a given feature. The features shown in bold are new in this work (see  \S\ref{sec:features}). 
%See text on initial values, mean value ($\bar{f}$), and normalization.
}}
\begin{footnotesize}
\begin{tabular}{llllll}
\toprule
% {\bf Class} & {\bf Feature} & {\bf Init. Val.} & {\bf Raw Range ($f$)} & {\bf Mean Val. ($\bar f$}) & {\bf Normalization} \\ 
{\bf Level} & {\bf Feature} & {\bf Init. Val.} & {\bf Raw Range} & $\bar x$ & {\bf Normalization} \\ 
\midrule
Packet%-level     
& Time-to-live of packet $p$                     & -       & [0, $T$]      & -     & $(x+1)/(T+1)$ \\                
\midrule 
Device%-level 
& Queue length at $v$                            & -       & [0, $B$]      & -     & $(x+1)/(B+1)$ \\
& Node density of $v$                            & -       & [0, $N$-1]    & -     & $\min \left(1, (x+1)/(N+1)\right)$            \\
& Node degree of $v$                             & -       & [0, $N$-1]    & -     & $\min \left(1, (x+1)/(S+1)\right)$ \\
& {\bf Dynamic connectivity (short)}, $\delta_s = 100s$  & -       & [0, $N$-1]    & -     & $(x+1)/N$     \\
& {\bf Dynamic connectivity (long)}, $\delta_l = 500s$    & -       & [0, $N$-1]    & -     & $(x+1)/N$     \\
& {\bf Dispersion (short)}, $\tau_s = 10s$    & $L/20$m & [0, $L$]      & -     & $(x+1)/(L+1)$ \\
& {\bf Dispersion (long)}, $\tau_l = 100s$    & $L/5$m  & [0, $L$]      & -     & $(x+1)/(L+1)$ \\
\midrule 
Path%-level 
& Queue length at $v$ for packets to $d$         & $0.1B$  & [0, $B$]      & -     & $(x+1)/(B+1)$ \\
& Euclidean distance                   & $L/2$m  & [0, $L$]      & -     & $(x+1)/(L+1)$ \\
& Transitive timer                               & 200s    & [0, $\infty$] & 200s  & $1/(1+e^{-k(x-\bar x)})$ \\
& {\bf AoI}                             & 200s    & [0, $\infty$] & 200s  & $1/(1+e^{-k (x-\bar x)})$ \\
\midrule 
Neighborhood%-
&  Aggregate (min, max, and mean) of device-level    & -  & -  &  -  & -  \\
%level 
& and path-level features of $v$'s neighbors & -  & -  &  -  & -  \\
\midrule 
Context%-level 
& Boolean flags: relative visit history for $p$ for action node  $u$  & -  & -  & -   & -         \\
 & Boolean flags: transmit vs. stay & -  & -  & -   & -         \\
\bottomrule
\end{tabular}
\end{footnotesize}
\label{tab:features}
\end{table*}

\subsection{Quality of Information Metrics}
\label{sec:qoi-features}
In a mobile network, nodes may be able to only infrequently exchange features, % control state,
decreasing the quality and timeliness of feature values.  
% But forwarding decisions can differ significantly depending on measured feature quality, for instance, whether the estimated value of a feature is believed to be close to the true value of the feature. For sparse networks, when forwarding decisions are most difficult, 
But %forwarding decisions can differ significantly depending on  the values of features, and 
without a way to measure feature quality, a node cannot distinguish whether its estimated value of a feature is close to or far from the true value of that feature. 
%For sparse networks, when forwarding decisions are most difficult, feature values are also typically most out-of-date and most uncertain with few opportunities to recover from bad forwarding decisions, making it doubly critical to characterize feature quality. 
For sparse networks, which have few opportunities to recover from bad forwarding decisions, feature values are also typically more out-of-date and uncertain than in dense networks, making it doubly critical to characterize quality. 
For example, in each setting in Table~\ref{tb:training-cases},
%to \ref{tb:use-cases}, %when the transmission range is 20m, 
feature quality %the quality of information 
can be significantly worse for a transmission range of 20m compared to 80m.
% when the transmission range is 20m than when it is 80m. 

Many %quality 
metrics can be used to represent quality of information. In addition, different %types of %quality 
metrics may be appropriate for different features.
%There are many possible quality metrics which could be considered, and for different features, different types of quality metrics may be appropriate. 
For instance, for features estimating a distribution, such as inter-meeting time, variance could be used, while for other features, measures of confidence or freshness may be more appropriate. Here, we focus on feature freshness as a metric as it is relatively easy to measure, unlike, for instance, variance which can require many samples for accuracy and so is hard to estimate in sparse networks.
%, and hence take a long time to obtain in sparse networks.
%more than a few samples of feature values with such samples not easily obtained in a sparse network when they are most useful. 

% \vspace{0.1in} \noindent 
{\em Age of information (AoI)\cite{yates2021age}.}
% We use AoI to characterize the freshness of features estimated from out-of-date information collected from other devices. An example of one such feature that we use in this paper is the estimated Euclidean distance from a device $v$ to some other device $d$ (see Table \ref{tab:features}). Each device maintains the last known location of every other device $d$ in the network along with a timer measuring the relative time $\Delta_d$ that has passed since that location was recorded; specifically, $\Delta_d$ starts at 0 and is incremented by one every timestep. When devices meet, they exchange locations  and timers for all devices, update their own estimated locations of devices to be those with the smallest timers, and increment each timer by one after each timestep.  The use of timers avoids the need for clock synchronization among devices. We then define the age of information for features associated with a device $d$ simply as $AoI_d=\Delta_d$. \bing{Explain the difference of 
We use AoI to characterize the freshness of features that are estimated using out-of-date information collected from other nodes. One such feature is Euclidean distance (see Table \ref{tab:features}). To estimate Euclidean distance, each node maintains the last known location of every other node $w$
% \tommy{I think we should use a parameter other than d here as d is used to represent destination} 
in the network along with a timer measuring the relative time $\Delta_w$ that has passed since that location was recorded; specifically, $\Delta_w$ starts at 0 and is incremented by one every timestep. When nodes meet, they exchange locations and timers for all nodes, update their own estimated  locations to be those with the smallest timers,
% \bing{Is the timer value copied as well?} \vicky{yes it is}
% \tommy{Timestamp is one of direct features so is copied directly from neighbors, after which the AOI is computed.}, 
and increment each timer by one every timestep.  The use of timers avoids the need for clock synchronization among devices. We then define AoI for those features associated with a node $w$ simply as $AoI_w=\Delta_w$. 

%\bing{Explain the difference of AoI and transitive timer.} 
While AoI appears to be similar to the transitive timer feature in Table \ref{tab:features}~\cite{manfredi2024learning}, 
% the main difference is that AoI directly measures the age of feature values, while the transitive timer stems from utility-based routing (see \cite{singlecopy-spyro}) which uses timer transitivity as an estimate of utility or goodness of a node as a possible next hop. 
% \tommy{While AoI appears to be similar to the transitive timer feature in Table\ref{tab:features}~\cite{manfredi2024learning}, 
the main difference is that AoI directly measures the age of %feature values, 
features (e.g., $w$'s location),
while the transitive timer, introduced in utility-based forwarding (see \cite{singlecopy-spyro}), is a  heuristic to estimate the goodness of a node as a possible next hop.  
Specifically, the timer transitivity calculation itself at a node $v$ combines the age of $w$'s location information (as in AoI) with the time needed to travel the distance between node $v$ and the neighboring node providing $w$'s location.

% We normalize $AoI_d$ by the time interval given in Table \ref{tab:features},  which approximates the time period for which features are expected to be good. 
%
%We note that while we explore the simplest version of AoI here, other variations are also possible, see \cite{yates2021age}. 

%%%%----------------------------------------------------------------%%%%%
\subsection{Network Memory Metrics}
\label{sec:memory-features}

Intuitively, larger AoI values should indicate that the associated feature values are 
% more out-of-date 
less accurate, but this is not necessarily always the case. %For instance, 
Consider a mobile network in which   %mobility model where 
nodes move in small loops. 
% move in a small neighborhood. 
In this case, even if nodes move very fast, the estimated feature values may stay close to the true values for a long period of time. In contrast, for other kinds of node movement
% in other mobility models 
(such as that given by the random waypoint model), estimated feature values (like the Euclidean distance to a destination) may quickly diverge from their true values. 
%depending on node movements, estimated feature values may stay close to the true values for long periods of time (e.g.,  for a very long time), while in other networks, estimated values may quickly diverge from true values. 
More generally, % for many measures of feature quality, 
some way to measure the rate of change in feature quality is 
needed. We call such measures {\em network memory metrics}: for instance, only in networks with short memory do large AoI values correlate with poor information quality. %; this is
%while this is not %it is not %necessarily 
% the case 
%unlike in networks with long memory.

\iffalse 
While Quality of Information metrics focus on directly quantifying the quality of specific features, network memory metrics focus on  quantifying network, neighborhood, and node movements \bing{what does `network movements' mean?} over time, since such movements themselves can impact feature quality. For instance, depending on node movements, estimated feature values may stay close to the true values for long periods of time, while in other networks, estimated values may quickly diverge from true values. More generally, for many measures of feature quality, some way to measure the rate of change in quality is also needed. We call such measures {\em network memory metrics}. 
\fi 

Mixing time, which measures how quickly a node's location becomes independent from its current location, is a natural measure to quantify network memory. The study in \cite{singlecopy-spyro} has characterized the asymptotic mixing time of RWP  and random walk mobility, to gain insight into the impact of mixing time on the performance of different forwarding strategies. Yet, while the mixing time of RWP  mobility can be  estimated from  average node speed and network area \cite{ferriere2003agematters}, estimating mixing time online for general node mobility %mobility models 
requires a large number of samples and computationally expensive statistical calculation.
%, which is impractical to do in a sparse mobile network. 
Consequently, we explore other ways to quantify network memory. 

Specifically, we quantify how quickly the quality of the estimated Euclidean distance feature decays as AoI increases, and use that as our network memory feature.
%
%In this work, as our network memory feature, we particularly focus on quantifying how quickly the quality of the estimated Euclidean distance feature decays as AoI increases. 
We can relate the rate of decay of feature quality
to the farthest distance away that 
% to how far 
a node has moved from a previous location within a given time interval. For instance, if a node stays near the same location for a long time period, we expect that the quality of feature estimates for that node will remain consistent for a long period of time. Previous work~\cite{ferriere2003agematters} defines the locality of a mobility process to be the time scale over which the position of a node's future location is correlated with its current location, and demonstrated empirically the locality of random walk and RWP 
mobility processes. This locality metric, however, is computationally intensive to estimate.
% difficult to be computed in an online manner. 
% In the following, 
We next propose a metric that we call {\em dispersion} that represents non-locality and can be estimated efficiently in an online distributed manner.

%is opposite to locality in use We define locality differently than in~\cite{ferriere2003agematters}  so that it can be estimated in an online distributed manner. Furthermore, we consider locality over two different timescales to better capture the rate of change of a device's locality. 

% \vspace{0.1in} \noindent
{\em Dispersion.}  We define dispersion as the farthest Euclidean distance that a node travels away from a starting point during some time interval $\tau$. When large distances are traveled by a node in a short period of time, then the node's movements should have a short mixing time, and the quality of estimated features regarding the node should degrade quickly as its AoI increases. Consider a node $v$ at location $(x_{t-\tau}, y_{t-\tau})$ at time $t-\tau$. Let $d_{t,\tau}$ be the maximum distance from $(x_{t-\tau}, y_{t-\tau})$ to any location that $v$ is at during the interval $[t-\tau, t]$. We take an exponentially weighted average of $d_{t,\tau}$ over intervals to obtain the dispersion feature $D_{\tau} = \beta d_{t, \tau}+ (1-\beta) D_{\tau}^{prev}$ 
that we use, 
% computed as 
% \begin{eqnarray}
    % $D_{\tau} = \beta d_{t, \tau}+ (1-\beta) D_{\tau}^{prev}$
% \end{eqnarray} 
% \noindent 
where $0 \leq \beta \leq 1$ and  $D_{\tau}^{prev}$ is the previous estimate of dispersion. 
To quantify the rate of change of dispersion, we  compute it for both  a {\em short interval}, $\tau_s$,  and a {\em long interval}, $\tau_l$.  We consider the relatively short timescales of $\tau_s=10$s and $\tau_l=100$s (see Table \ref{tab:features}).
in our evaluation in \S\ref{sec:eval}.
%, as long distances traveled in a short period correlate with fast mixing, compared to long distances traveled in a long period \bing{What does this sentence mean?}\bing{Do we expect faster nodes to have larger locality? Locality is also related to mobility models; even fast nodes in some mobility models can have low locality}. \bing{locality of node movement: Do I wander around in a local region; or I have moved far away. } 

%%%%----------------------------------------------------------------%%%%%
\subsection{Community Metrics}
\label{sec:community-features}
We refer to metrics that quantify the variety and rate  of neighborhood changes as {\em community} metrics. 
In mobile networks, the neighborhood changes observed by nodes can be affected by many factors, including node speeds, movement patterns, and %the specific
spatial distribution. % of nodes. 
The resulting changes in node neighborhoods can have a significant impact on forwarding decisions. For example, nodes that move faster than others will also meet other nodes more frequently, and hence might have a higher chance of quickly delivering a packet to its destination, and so could be favored as a packet's next hop. However, in a densely connected network, even a node moving at a low speed might meet other devices fairly frequently, and hence, %that node, 
rather than forwarding its packets to fast nodes, a node could simply wait until it meets a packet's destination. Doing so could avoid %incurring additional 
unnecessary forwards without increasing delay.
% incurring long delay.

%Varying device speeds and movement patterns and the specific spatial distribution of devices all impact the variation in neighborhood changes seen by devices, which then impacts forwarding decisions. For example,  devices that frequently meet other devices have a higher chance of delivering a packet to its destination and so might be favored as a packet's next hop.  if a device frequently meets other devices, then that device could simply wait until it meets a packet's destination, rather than incurring additional unnecessary forwards. We refer to metrics that quantify the variety and rate  of neighborhood changes as {\em community} metrics. 

% \vspace{0.1in} \noindent
{\em Dynamic connectivity.}
% \vicky{Dynamic connectivity extended to two timescales. Explain why the two timescales.}
% We propose the {\em rate of connectivity}
% propose one feature 
%focus on 
To quantify a node's community, %the rate at which nodes meet.
% which gives  information about the fraction of nodes in the network seen \bing{Remove ``which gives  information about the fraction of nodes in the network seen''?}. 
%To do so, 
% The rate of connectivity extends 
we extend the {\em dynamic connectivity} metric in \cite{multicopy-spyro}, which quantifies the number of unique nodes encountered by a given node  within a time interval, to consider two
%multiple 
timescales (short and long) %to quantify changes in connectivity,
and  design an online distributed approach for its estimation. %to estimate dynamic connectivity.  
%Specifically, 
The use  of two timescales allows us to obtain the rate of change of dynamic connectivity.
To compute dynamic connectivity, a node $v$ first records
%consider a device, $v$, that records 
the most recent time it met every other node in the network. Then at time $t$, node $v$  counts the number of unique nodes that it saw within a short time interval  $[t-\delta_{s},t]$ and a long time interval $[t-\delta_{l},t]$ to compute dynamic %the rate of change of
connectivity at two timescales. 
%{\em short dynamic connectivity} and {\em long dynamic connectivity}, respectively. 
In our evaluation, 
 we set $\delta_{s}=100$s and $\delta_{l}=500$s (see Table \ref{tab:features}).  
% For dynamic connectivity at both short and long timescales, we normalize them with respect to the total number of devices in the network, as in Table \ref{tab:features}, to obtain a ratio \bing{This sentence is about normalization. Move it to later?}.

\subsection{Feature Initialization and Normalization} \label{sec:feature-norm}

% In Table \ref{tab:features}, the initial values are needed to compute path features for devices that have not yet been met and to compute features that are a time-weighted average (e.g., for locality features) \bing{I see we also need initial values for some device features?}. 

Our goal with feature initialization 
%in Table \ref{tab:features} 
is to ensure that a feature always has a default value even when the information needed to estimate that feature is not available.  This is specifically the case for time-averaged features (i.e., dispersion) and path features (which inherently describe a relationship between two nodes and so lack a value until those nodes meet or third-party information is shared). 
We initialize dispersion to $L=1500$m, 
comparable to the size of the large networks that we evaluate.
% which is approximately half the size of the largest network area we evaluate in \S\ref{sec:evaluation}\bing{This sentence is a bit strange: the unit of disperse is meter and the unit of area is square meters. What about `We initialize dispersion to $L=1500$, comparable to the size of the large networks that we evaluate'.}.
We initialize the path features to be close to the worst-case values we see in cumulative distribution functions (CDF) plots of feature values, to be conservative. For instance, we initialize destination queue length to a fraction of the maximum buffer size $B=200$ in training (with $B=2000$ in testing to ensure that no packets are dropped due to buffer overflow),
%\bing{Is it 2000?} 
and Euclidean distance to $L/2$.
We initialize the temporal features of transitive timer and AoI to $200$s, %\bing{$AoI$ or just AoI?}, 
which approximates the time period for which features are expected to be good. 

Our goal with normalization  in Table \ref{tab:features} is to ensure that all features are relatively similar in terms of their scale of values to make DNN training easier. For features for which the maximum value is clear, that value can be directly used for normalization,
such as time-to-live being bounded by the initial time-to-live of $T=300$ in training (and $T=3000$ in testing to avoid drops), rate of connectivity being bounded by the number of nodes in the network $N$,
queue length being bounded by $B$, and
dispersion being bounded by $L$.
Node density we normalize by $N$, to capture the fraction of nodes in the network to which a node is connected, while node degree we normalize by a scaling factor of $S=10$, to capture the point at which a network is unequivocally well-connected.
For features that are unbounded, specifically the transitive timer and AoI, we use a shifted sigmoid function, $\sigma_k(x-\bar{x})=\frac{1}{1+\exp(-k(x-\bar{x}))}$ to force an upper bound on values. This is a non-linear mapping robust against outliers.
The mean value, $\bar{x}=200$, is set empirically,
% based on estimating the mean from CDF plots of the feature values
% \bing{Is $\bar{x}$ indeed the mean value? I assume the mean can vary significantly based on transmission range? Maybe just say that we set $\bar{x}$ to 200s empirically, without saying that it is the mean.}, 
while  $k=0.01$ is a scaling factor varying how sharply values approach their upper limit.

\section{Continual Learning Based DRL Forwarding} 
\label{sec:drl_forwarding}

%\subsection{The Need for Continual Learning}

%wb, 7/17/2025, only a small figure for space
\tydubfigsingle{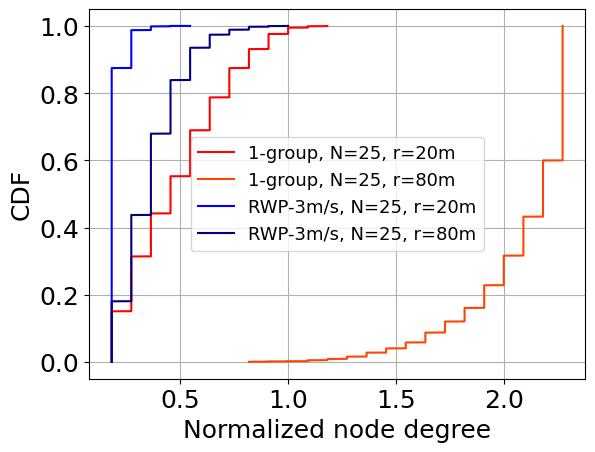}{}{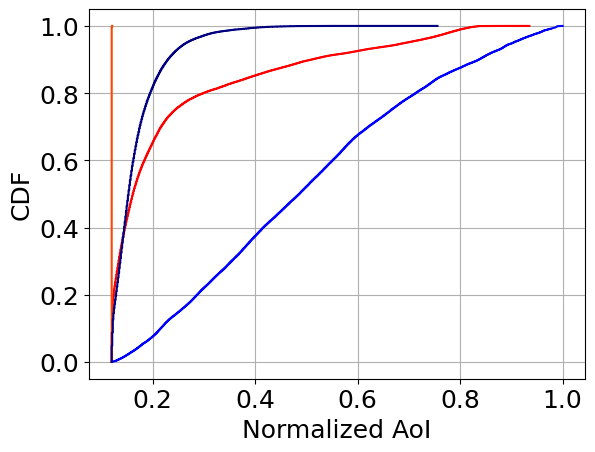}{}{Distributions of two feature across four network scenarios. 
}{features}

\subsection{Methodology}
% Network 
Node diversity can give rise to drastically different kinds of network connectivity.
% Diverse networks can have drastically different characteristics.
% One example is in Fig.~\ref{fig:features}, which 
For example, Fig.~\ref{fig:features}
shows %that
the CDFs
%the cumulative distribution functions (CDFs) 
of two features. We see that the features differ significantly across the four network scenarios, sometimes with little overlap in distribution. 
%wb, 7/24/2025, already mentioned earlier
\iffalse 
To train DRL forwarding strategies across diverse network scenarios, a simple strategy is to train a model on one scenario, then input the model to train on the next scenario, and so on. This approach, however, 
%is known to 
suffers from ``catastrophic forgetting''~\cite{McCloskey1989}, i.e., models trained on later scenarios do not perform well on earlier scenarios. 
\fi 
% \vicky{The following sentence comes across a bit out of place given the previous.}
CL addresses this by 
%overcomes the above challenge and 
% allowing 
learning across a sequence of scenarios, retaining 
early knowledge from training, 
while incurring bounded computational resources and memory.
%an intelligent system to retaining the old knowledge while relying on bounded computational resources and memory footprint
%adapt to diverse scenarios without ``catastrophic forgetting''. 

While many techniques have been proposed for CL, we focus on {\em experience replay}~\cite{Ratcliff90:ER} coupled with {\em balanced reservoir sampling}~\cite{Vitter1985}. This approach is simple and robust, and has been shown to be as effective as more sophisticated approaches~\cite{Buzzega20:bag}. The main idea of experience replay is to interleave past examples and current training examples. The approach of balanced reservoir sampling uses similar amounts of samples from each scenario so that the trained model is not biased toward 
any one scenario.

As in~\cite{manfredi2022learning,manfredi2024learning}, we use {\em centralized offline training} to train a DRL-based forwarding model; after training, each node uses the trained DRL model to make forwarding decisions  independently in a {\em distributed online} manner. Offline training allows gathering of data from all nodes (which is not feasible 
in online training), 
% during the offline training phase,
%of centralized information (all data from all nodes) 
leading to more effective training.
% when training in an online manner.    
% Specifically, 
To this end, we develop a network simulator to both generate training data and train the DRL model through CL.

%We next describe our continual learning approach.
%of continual learning for  $n$ scenarios, $n\ge 2$. 
For CL, %we start with training 
we first train a DRL agent in the first scenario as in~\cite{manfredi2022learning,manfredi2024learning}, which do training in {\em rounds}.
%, leading to dataset $D_1$. 
% Specifically, the training happens in {\em rounds}. %(1000 timesteps). 
In each round, the current DRL model is used to make forwarding decisions, and the data collected so far is used for training at the end of the round. 
%The data is organized in state, action, reward, and next state tuples, $(s, a,r,s')$, where state includes all the state and action features in Table~\ref{tab:features}.
%wb, 7/24/2025
%, action records the action that was selected, and reward is as defined in \S\ref{sec:bk-DRL-forwarding}. Let $D_1$ denote the dataset, in the form of  $(s, a,r,s')$, that is used to train the model for the first scenario.  % 
But unlike in~\cite{manfredi2022learning,manfredi2024learning}, after we complete training in the first scenario, 
% completing training for the first scenario, 
we proceed to train in the second scenario, and so on.
%Specifically, 
Consider training for round $t$ in the $k$-th scenario, $k\ge 2$.
Let $D_k(t)$ denote the data %that has been 
gathered so far in the $k$-th scenario. Let $D_1,\ldots,D_{k-1}$ denote the data gathered from the first $k-1$ scenarios. 
%In this case, we use the data, $D_k$, 
% that is being 
%gathered in this scenario together with the data from the first $k-1$ scenarios, i.e., $D_1,\ldots,D_{k-1}$. %All the 
Each dataset is organized into state, action, reward, and next state tuples, $(s, a,r,s')$, 
where the corresponding state and action features include all of the
% where state includes all of the state and action 
features in Table~\ref{tab:features}. To be computationally efficient, we use sampling, i.e., we take a fraction, $\alpha$, of data from each dataset. For ease of exposition, 
%assume that the datasets, $D_1,\ldots,D_{k-1}$, are sufficiently large, while 
assume $D_k(t)$ is the smallest dataset (since it is the current dataset being gathered, growing through round $t$ and initially empty in the first round).
% ; it grows with round $t$ and starts with a small size).
%wb, 7/24/2025, for space
%\footnote{If this is not the case, we can extend the following method in a straightforward manner to consider the dataset that is the smallest.}.  
Let $n=\max(\alpha|D_1|, \ldots, \alpha|D_{k-1}|, \alpha|D_k(t)|)$. If $n \ge |D_k(t)|$, i.e., there 
% is not enough data 
is insufficient data
in $D_k(t)$ even if all samples collected so far are included, we set $n$ to $|D_k(t)|$. 
% After that, 
Then, we choose $n$ samples uniformly randomly from each dataset, $D_1,\ldots,D_k(t)$. The $n$ samples in each dataset are divided into batches, each of size $b$.
% \vicky{this sounds like we just use the data from the t-th round for each scenario? Is this true? Do we not sample earlier rounds? Also should it be $D_1(t),\ldots,D_k(t)$ with a t on $D_1$?}
Finally,
% After that, 
a batch is taken uniformly randomly from the $k$ datasets 
% at each point 
until all batches have been used for training.

%wb, 7/24/2025, not sure whether to include this or not; comment out for now for space
%
%\bing{Todo.} For each scenario, we set the initial DNN model to randomized weights, which tend to lead to better results than set it to the previous model.   
   
To summarize, the above approach uses the same amount of samples from each dataset, and mixes the batches from the $k$ datasets uniformly randomly for training. 
%Without otherwise stated, 
In the rest of this paper, we use $\alpha=10\%$ and $b=32$.
%since we found $10\%$ sampling is sufficient and a batch size of 32 leads to good results. 
We also tried other options and found that using $\alpha=5\%$ leads to worse results,
%than $\alpha=10\%$,
while using $\alpha=30\%$ leads to similar results as $\alpha=10\%$; using $b=64$ leads to similar results as $b=32$.

\begin{table}[t]
    \centering
    \caption{
    The full 
    space of possible training scenarios 
    considered for CL.
    %in the design of our continual learning agent. 
    The three settings 
    in bold are the only ones used for CL. 
    % Those settings shown in bold are the only scenarios that we use.
    % ultimately train on and use in continual learning. 
    % 5Note that all scenarios are for small networks (24 or 25 nodes moving in 500 m $\times$ 500 m area).
    %, with average node degree for each transmission range ($r=20$, 50 or 80m) shown.
    % In mixed settings, the two numbers indicate the counts of slow and fast nodes, respectively.}
    }
    \begin{tabular}{lc|ccc} % Retained the | for the vertical line
        \hline % <--- Replaced \toprule
        \multirow{2}{*}[-1.5pt]{\textbf{Setting}} & \multirow{2}{*}[-1.5pt]{\textbf{$\bar v$ (m/s)}} & \multicolumn{3}{c}{\textbf{Average node degree}} \\
        \cline{3-5} % <--- Replaced \cmidrule and adjusted span
        & & $r=20\text{m}$ & $r=50\text{m}$ & $r=80\text{m}$ \\
        \hline % <--- Replaced \midrule
        RPGM 1-group & 3.0 & 3.2 & \fbox{\textbf{13.6}} & 21.1 \\
        \hline % <--- Replaced \midrule
        RPGM 2-group & 3.0 & 1.8 & 7.2 & 11.1 \\
        \hline % <--- Replaced \midrule
        RWP-3m/s & 3.0 & 0.2 & \fbox{\textbf{1.2}} & 2.7 \\
        RWP-15m/s & 15.0 & 0.2 & 1.0 & 2.5 \\
        RWP-mix1 (5,20) & 12.6 & 0.2 & 1.0 & 2.5 \\
        RWP-mix2 (12,13) & 9.2 & \fbox{\textbf{0.2}} & 1.1 & 2.7 \\
        RWP-mix3 (20,5) & 5.4 & 0.2 & 1.1 & 2.7 \\
        \hline % <--- Replaced \bottomrule
    \end{tabular} \label{tb:training-cases}
    \parbox{\linewidth}
    {\footnotesize *RPGM 2-group contains 24 nodes (12 nodes in each group), all others  contain 25 nodes,
    %All scenarios contain 25 nodes (except for RPGM 2-group, which contain 12 nodes in each group) 
    moving in 500 m $\times$ 500 m area. In mix settings, the \#s of slow and fast nodes are marked (in order) in the parentheses.
    %*In mixed settings, the two \#s are the counts of slow and fast nodes, respectively.
    }
\end{table}

\subsection{Choosing Scenarios for Continual Learning } \label{sec:choose-scenario-CL}
With the above CL  algorithm, we can train a DRL agent using many scenarios sequentially. The more scenarios %that are being 
used, in principle, the more generalizable the trained model. However, training over many scenarios is time consuming and computationally intensive. Therefore, a guideline for CL is to carefully choose the training scenarios used to limit training time~\cite{Wang2024:CL}.
% to limit the number of scenarios

%
%so that a model trained using only a small number of scenarios is %already sufficiently generalizable~\cite{}.  Our goal in this work is to train a generalizable forwarding strategy that can be used for real-world scenarios.

{\em Space of possible training scenarios.} 
% Our goal is to train a generalizable forwarding strategy that can be used for real-world  %routing 
% scenarios. 
We consider a total of $7 \times 3=21$ possible training scenarios as shown in Table~\ref{tb:training-cases}. 
These scenarios include Reference Point Group Mobility (RPGM)~\cite{hong1999group} with one or two groups, and Steady-state Random Waypoint (RWP) mobility  \cite{navidi2004stationary,le2005perfect}. %with homogeneous or heterogeneous (mix of fast and slow) node speeds. 
A node moves either {\em slow} (from 1 to 5m/s with mean of 3m/s) or {\em fast} (from 13 to 17m/s with mean of 15m/s). In RPGM 1-group and 2-group scenarios, all nodes are slow. For RWP mobility, we consider two homogeneous settings (all nodes are slow or fast) and three heterogeneous settings, RWP-mix1, RWP-mix2, and RWP-mix3, with the number of slow and fast nodes marked in Table~\ref{tb:training-cases}. 
%For one mobility model, 
The transmission range $r$ varies from 20m to 80m, resulting in
networks ranging from poorly- to well-connected.
% leading from poorly-connected to well-connected networks.
%, with the average node degree from 0.2 to 21.1.
All mobility 
traces are generated using BonnMotion \cite{aschenbruck2010bonnmotion}.

In each scenario, we generate time-varying traffic. 
% by varying flow and packet arrivals, and flow durations, over time. 
Flows arrive following a Poisson distribution with parameter $.001 N / 25$, where $N$ is the number of nodes in the network. Packets on a given flow arrive according to a Poisson distribution with parameter 0.01. Flow durations follow an exponential distribution with parameter 5000. 
% The buffer at each node is set to 2000 packets \bing{Is this related to $B=200$ that was described earlier? Should $B$ in Section IV D be 2000?} so that no packets are dropped due to buffer overflow. 
We further assume sufficient bandwidth for
% so that 
all packets at a device to be transmitted in a single timestep.
%Initially, there are 50 flows in the network, which corresponds to the average number of flows given flow arrivals and durations. We set the maximum length of queues to 2000 packets, which is sufficiently large that no packets are dropped due to buffer overflow. We assume sufficient bandwidth for all packets at a device to be transmitted in a single timestep:
 %for the simulation results reported in this paper:  this is because in sparse networks  devices generally have few if any neighbors with which to share bandwidth and  minimal interference of transmissions. 

{\em  Scenarios selected for CL.} During CL, we start with the simplest mobility scenario, RPGM 1-group, with $r=50$m. %After that, 
Then the second scenario selected is homogeneous RWP (specifically, RWP-3m/s, $r=50$m), and %then
the third scenario is heterogeneous RWP (specifically, RWP-mix2 which has a similar number of fast and slow nodes with $r=20$m.
% , where we use RWP-mix2 which has a similar number of fast and slow nodes with $r=20$m. 
These three scenarios, marked in bold in Table~\ref{tb:training-cases}, are for small networks (25 nodes %moving 
in a 500m$\times$500m area) so that training is computationally efficient.
%; the testing scenarios in \S\ref{sec:results} and \S\ref{sec:real-world} are much larger to test the  scalability and generalizability of our approach.
As shown 
in Table~\ref{tb:training-cases}, 
%the scenarios
%these three scenarios 
they cover a wide range of settings, including three mobility settings, with the average node degree ranging from 0.2 to 13.6. 
%
%As such, we conjecture that a DRL forwarding policy continually trained by them can be sufficiently generalizable.

Although we only use three scenarios for CL, 
each scenario includes a large number of timesteps: 100K, with the last 40K timesteps as cool-down period\footnote{
No new packets are generated in cool-down period, so that all packets that have been generated are delivered by the end of the simulation.},  
 each timestep corresponding to one second. This allows us 
to fully capture the variability in each scenario. 
 %Specifically, we use  timesteps for each scenario,
%corresponds to one second, 
%with the last 40K timesteps as cool-down period so that no new packets are generated, and packets that have been generated are delivered by the end of the simulation. 
Note that for each scenario, training happens at the end of each round of 1K timesteps. %; each round includes 1K timesteps. 
%We use a large number of timesteps for each scenario in training to capture the variabilities in mobile wireless networks.  

{\em Training environment and parameters.} We perform our simulations using a custom discrete-time  packet-level 
%network 
simulator we wrote in Python3, and used version 2.13.1 of Keras \cite{chollet2015keras} and Tensorflow  \cite{tensorflow2015-whitepaper}.
%to implement the  DNN. 
The DNN has $m$ feature inputs ($m=63$, see Table~\ref{tab:features}), two fully connected hidden layers (with $10m$ and $m/2$ nodes), 
and one output corresponding to the Q-value.
%for the input  features. 
We use the same reward function as in~\cite{manfredi2024learning}: $r_{stay} = -1$ if the packet stays at its current node,  $r_{transmit} = -2$ if the packet  moves to a neighboring node,  $r_{delivery} = 0$ if the packet reaches its destination, and  $r_{drop} = r_{transmit} / (1-\gamma)$ if the packet is dropped, corresponding to receiving  $r_{transmit}$ for infinite time steps where $0 \leq \gamma \leq 1$ is the RL discount rate, and is set to 0.99.
We use the  ReLU activation function,  Adam optimizer, mean squared error (MSE) loss, batch size of 32, 10 epochs, learning rate of $10^{-4}$, and training validation split of 0.2. 
%wb, 7/23/2025, decide later
For the RL agent,  we use an exploration rate, $\epsilon$, of 0.1.
%and a discount factor, $\gamma$ of 0.99. 

%%For all layers, the activiation function is ReLU, and the optimizer is Adam. We use mean squared error as the loss function. The training in each round uses learning rate \bing{$10^{-4}$??}, with batch size of 32, number of epochs of \bing{10??}, and training validation split of 0.2. 

%Consider the continual learning over the three scenarios, we refer the first model 

%{\em Testing scenarios.} 
After completing CL, we verify that the model trained in the three scenarios is indeed generalizable to the 21 scenarios in Table~\ref{tb:training-cases} (figures omitted).
%including RPGM 2-group mobility, even though it was not used in CL.
Other testing results are detailed in \S\ref{sec:evaluation} and \S\ref{sec:real-world}.
%In \S\ref{sec:evaluation}, we evaluate the model in 36 large mobile network scenarios with new mobility scenarios, and in \S\ref{sec:real-world}, we evaluate it in real-world scenarios.  

%Even more remarkable, as we shall see in \S\ref{sec:evaluation}, they are generalizable to 36 large mobile network scenarios with new mobility scenarios, with average node degree to 0.1 to 15.8. Furthermore, with fine-tuning in one real-world scenario in one city, the model is generalizable to 18 diverse real-world scenarios, even in another city (see \S\ref{sec:real-world}).        

%Choose three diverse scenarios (connectivity, mixture of speed) carefully. Consideration of realworld cases.  Computational overhead; 
%\section{Large-scale Network  Evaluation}
\section{ Evaluation in Large Networks}\label{sec:eval}
\label{sec:evaluation}
 
%wb, 7/24/2025, for space
%
%We now evaluate the DRL-based forwarding policy learned using CL in a wide range of large-scale testing scenarios.  

%all packets are delivered by the end of the simulation. 
%We perform our simulations using a custom discrete-time  packet-level network simulator that we have written in Python3, and use  version 2.13.1 of Keras \cite{chollet2015keras} and Tensorflow  \cite{tensorflow2015-whitepaper} to implement the  DNN. 

% \begin{figure}[t]%left, bottom, right, top
% \centerline{\hbox{
% \subfigure[ \textsc{rwp-test}]
%     {\includegraphics[width=1.5in, trim = 0.5cm 0cm 2.5cm 0cm, clip]{plots-new-features/scenario_file_nodes_trajectory_rwp1region.pdf}}    
% }}
% \vspace{-0.2in}
% \caption{{\small Example device mobility. We show the first 20 BonnMotion waypoints for the device IDs indicated.}}
% \vspace{-0.1in}
% \label{fig:mobility}
% \end{figure}

%%%%%----------------------------------------------------------------%%%%%
\mysubsection{Evaluation Setup}
We consider 36 test scenarios, each containing 100 nodes moving in a 1000m$\times$1000m area. The first 21 scenarios are the same as those considered for training  (Table~\ref{tb:training-cases}), except that they are $4\times$ larger. The last 5$\times$3=15 scenarios (5 %settings with different 
speed configurations and 3 transmission ranges, consistent with RWP) are for Manhattan Grid mobility~\cite{aschenbruck2010bonnmotion}.
In this model, the %1000m$\times$1000m 
area is divided into 20$\times$20 grid blocks (each block of size 50m$\times$50m); nodes move along the grid lines, and can only change direction at the four corners of a block. 
%Each test scenario includes 50K timesteps (with the last 20K as a cool-down period).
%
% grids (each grid block of size 50m$\times$50m); nodes move along the grid lines, and can only change direction at the four corners of a grid. 

We compare our CL-trained DRL model, {\em DRL-CL}, with three other strategies. (i) {\em Oracle strategy}, which provides the best performance, but not achievable in practice due to the lack of oracle information. We obtain the results of this strategy through  epidemic-based flooding~\cite{vahdat00:epidemic}, which leads to optimal delay when ignoring %not considering 
congestion.
%caused by packet flooding in the network. 
Specifically, we obtain the delivery delay for a packet as the delay when the first copy reaches the destination, and record the resource usage as the number of forwards of this copy. (ii) Two state-of-the-art heuristics, {\em Seek-and-focus} and {\em Utility}-based forwarding~\cite{singlecopy-spyro}. 
%We only present the results for Utility-based strategy below in the interest of space (it outperforms Seek-and-focus in heterogeneous scenarios). 
%
%We found Utility outperforms Seek-and-focus in heterogeneous scenarios, and only present the results for Utility below in the interest of space. 
%We therefore omit the results for Seek-and-focus below in the interest of space. 
%
%wb, 7/24/2025, will not be able to describe them clearly; just cite the original paper
%
%In Utility-based strategy,  each device maintains timers measuring how long ago it met with other devices in the network; when two devices $u$ and $v$ meet, all timers for all devices are exchanged.  If one of the devices, for instance $u$, has more recently met a device $d$, the other device, $v$,  will update its own timer estimate for $d$ based on $u$'s timer for $d$.
%using timer transitivity to be that of $u$'s. 
%A utility threshold is then used to decide whether to forward; we set this threshold to 10s empirically \bing{Vicky: can you explain this utility threshold more clearly? Is it after or before the timer exchange? Or maybe we should save some space not describing it and only cite~\cite{singlecopy-spyro}.}.

The performance metrics include {\em delivery rate}, {\em delivery delay}, and {\em number of forwards}. They quantify the percentage of the packets delivered, 
%an important metric, 
and the delay and number of forwards for  delivered packets. The latter two metrics  present tradeoffs between performance and the amount of resources consumed.

\tydubfigsingle{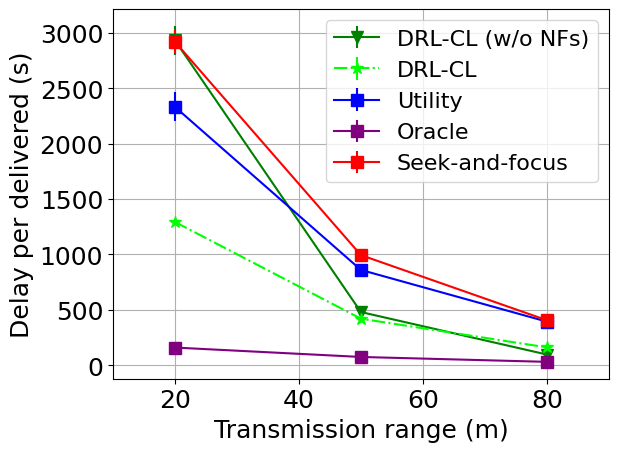}{Delay.}{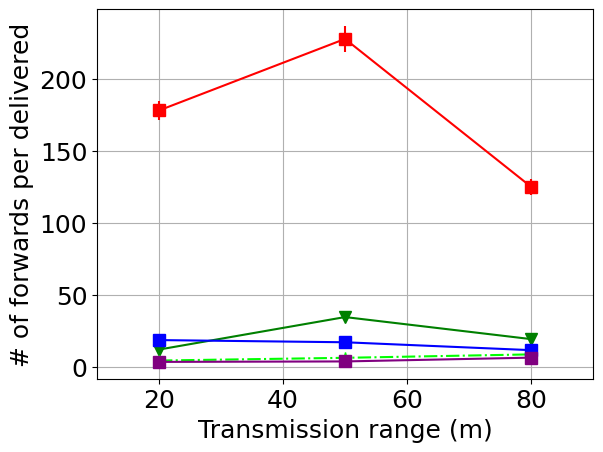}{\# of forwards.}{Comparing DRL-CL with and without new features (NFs) and three other schemes,
%and `DRL-CL w/o NF' (i.e., without new features), 
RWP-mix3 setting, large network.}{new-features-eval}

\tydubfigsingle{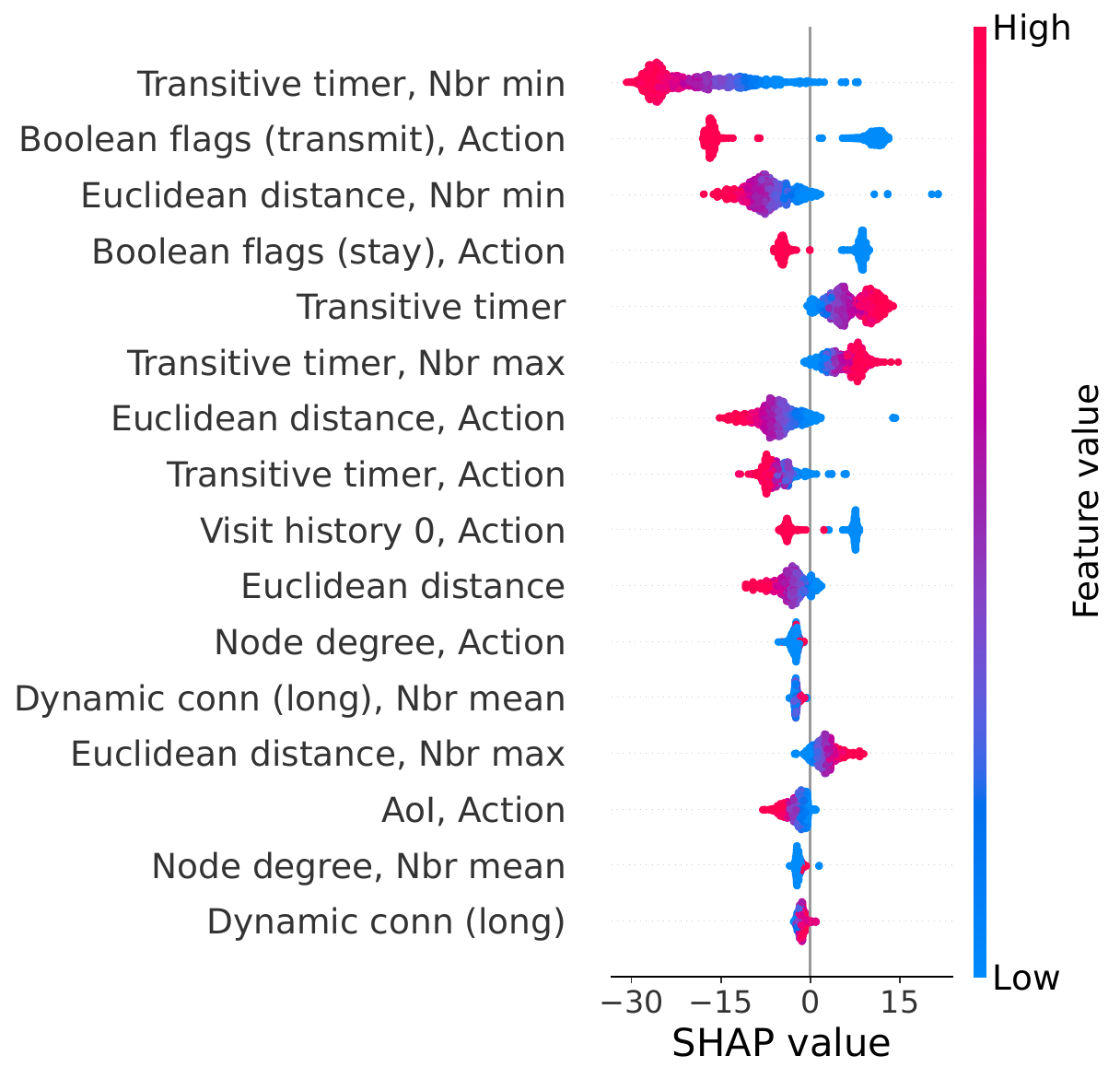}{$r=20\,\text{m}$.}{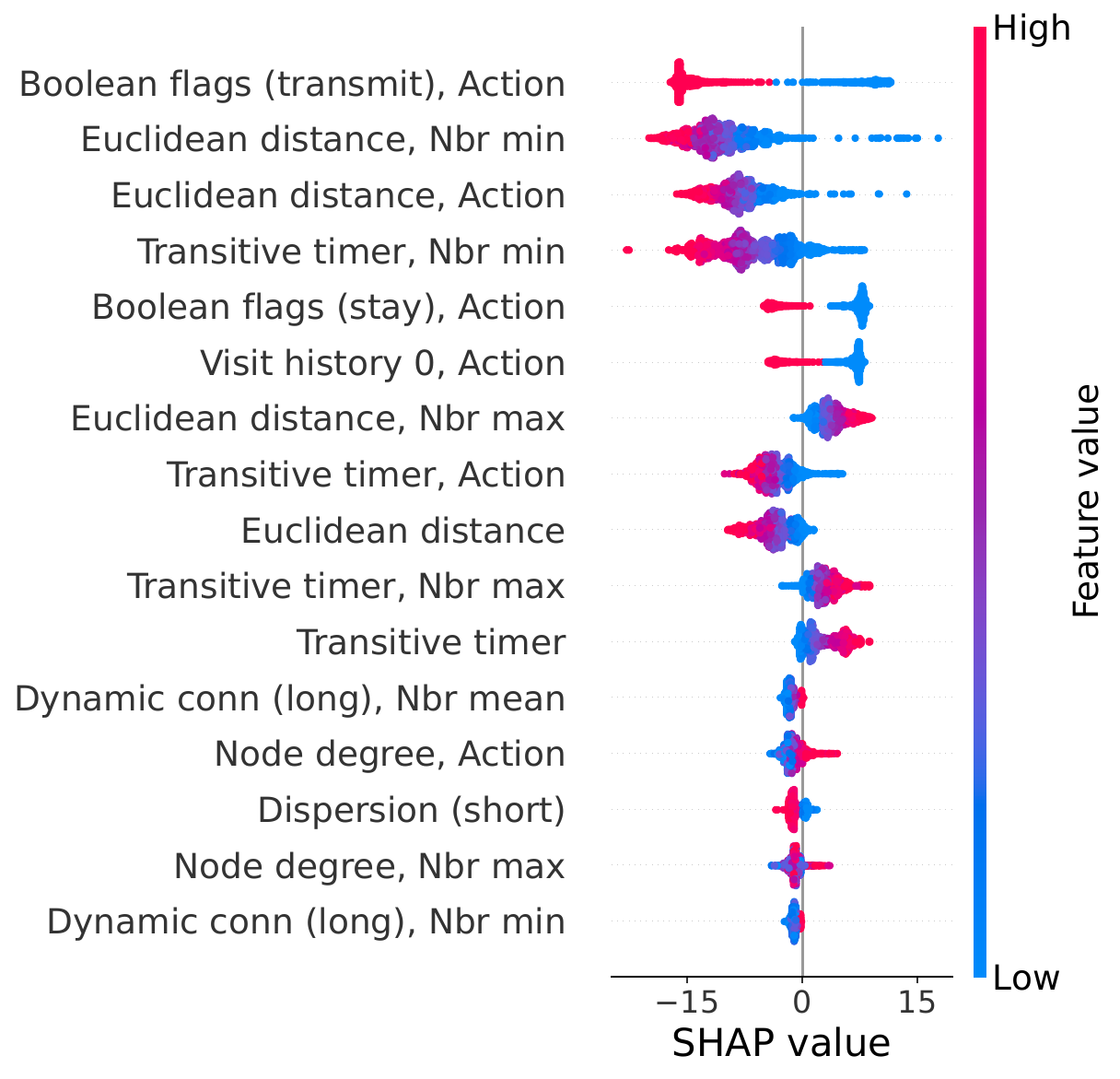}{$r=80\,\text{m}$.}{Shapley values of the top-16 important features for DRL-CL, RWP-mix3 setting, large network.  %\bing{Todo: decide the names of the features. Use names in Table I: Node density, Node degree, `Node density, Nbr mean', `Dynamic connectivity (long), Nbr min', `Visit history, Action'} 
}{Shapley}

\subsection{Evaluation Results} \label{sec:results}

%{\em The need for new features.} 
We find that the new features  (see \S\ref{sec:features}) are indeed helpful in capturing important network characteristics, leading to more generalizable models. As an example, we show the results for RWP-mix3 when the
transmission range $r$ varies from 20 to 80m. Fig.~\ref{fig:new-features-eval} shows the delay and number of forwards; all strategies lead to 100\% delivery rate (omitted in figure).
%Fig.~\ref{fig:new-features-eval} shows one example. It is for RWP-mix3. 
%where we compare DRL-CL with and without the new features. 
%(i.e., the policy trained with the new features) and that without using the new features. 
%For comparison, we further plot the results for Oracle and Utility-based strategies. 
%Only delay and number of forwards are shown in the figure, 
%since all strategies lead to 100\% delivery rate. 
%Compared to the model without new features, 
The model with new features significantly outperforms the one without new features: it leads to $57\%$ lower delay when the %transmission range 
$r=20$m, and 
many fewer forwards overall. It also leads to 
$45\%$ lower delay than the Utility-based scheme when $r=20$m and less forwards. Seek-and-focus leads to high delay and significantly more forwards than do the other schemes.
Fig.~\ref{fig:Shapley} shows the top-16 important features obtained by SHAP~\cite{NIPS2017_7062} when the transmission range is 20 or 80m. We see that 
which features are important
% the importance of the features 
differ across the scenarios; for both, %in both cases, 
we see our proposed new features (AoI, dispersion, and dynamic connectivity) among the top-16 features.    

%Describe much better results when using the new features, versus not using these new features. Also, show intuitively that model behaviors make more sense. Utility-based routing uses timer transitivity feature alone and does not generalize well. 

%\bing{7/16/2025. Compare the results with and without new features to show that the new features are helpful for generalization.} Show the results for one scenario CL2 with and without new features. 
%RWP-mix2-50-50 large network (actually will Use RWP-mix3 (80-20); four lines. Use Shapley values to show that some new features are in top-20. 

%\bing{7/16/2025. Compare the results to show the benefits of CL in three scenarios.} Compare 1-group, CL1 and CL2 (show that CL2 is better than CL1). Three curves + utility + oracle.    

\tydubfigsingle{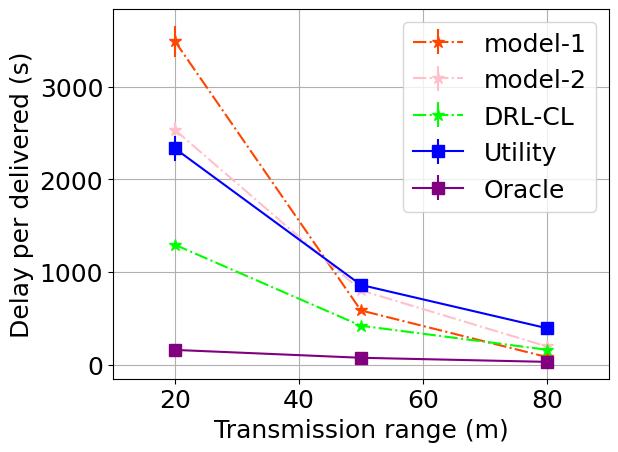}{Delay.}{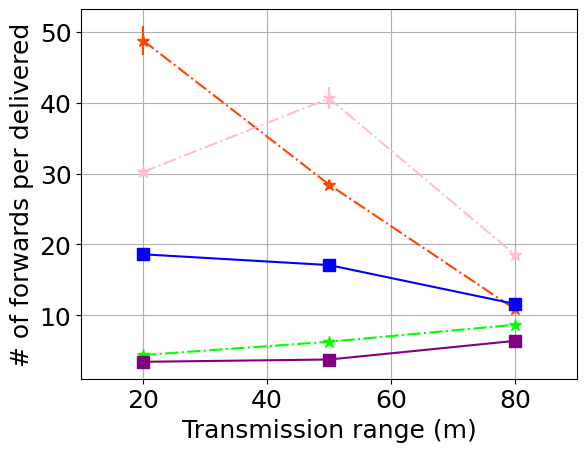}{\# of forwards.}{Comparing DRL-CL with models after training in first two scenarios (marked as `model-1' and `model-2'), RWP-mix3 setting, large network. 
}{CL3-CL2-CL1}
%%\bing{Add another homogeneous case; no space.} 

%{\em Benefits of CL.} As mentioned in \S\ref{}, we use CL in three scenarios. 
We further find that the CL-trained model on three scenarios (see \S\ref{sec:choose-scenario-CL}) leads to a more generalizable model than models trained only in one or two scenarios.  Fig.~\ref{fig:CL3-CL2-CL1} compares the results of DRL-CL with those of `model-1' and `model-2' (i.e., the models at the end of the first or second scenario in CL). %We see that
DRL-CL significantly outperforms the other two models when the transmission range is 20 or 50m. This is perhaps not surprising since the third scenario used in DRL-CL is for a transmission range of 20m with mixed speed. We %further
confirm that DRL-CL performs better for other scenarios (figures omitted).    

{\em Overall results.} Table~\ref{tb:testing-large-res} lists the delay and number of forwards for all 36 test scenarios. The three results listed in each cell are the average results for Oracle, DRL-CL, and Utility-based forwarding, respectively; 
the results for Seek-and-focus are omitted since it generally performs worse than Utility-based forwarding. 
%We only present the results for Utility-based strategy below in the interest of space (it outperforms Seek-and-focus in heterogeneous scenarios). 
For each setting, we run 5 simulations (each includes 50K timesteps, with the last 20K as the cool-down period) and confirm that the 95\% confidence intervals are tight.
%to obtain the mean and 95\% confidence intervals, and confirm that the confidence intervals are tight.  
We only show the average delay and number of forwards since the delivery rate is 100\% for all strategies, except for one case (Grid-3m/s, $r=20$m), where DRL-CL and Utility-based forwarding have delivery rate of 96\% and 95\%, respectively.
%does not fully deliver all the packets. 

%{\em Fine-tuning base model.} 
As expected, we see from Table~\ref{tb:testing-large-res} that
% the performance of 
DRL-CL %is 
performs worse
than the Oracle scheme for all cases.
Conversely, DRL-CL generally outperforms Utility-based forwarding. It only exhibits higher delay in one case (1-group mobility, $r=80$m). However, in this case, both delay values are relatively low (61s for DRL-CL vs. 33s for Utility-based forwarding).
%
%On the other hand, DRL-CL outperforms Utility-based forwarding in most cases. In only one case (1-group mobility, $r=80$m), DRL-CL leads to higher delay than Utility-based forwarding. On the other hand, in this case, both delay values are already fairly low (61 and 33s, respectively).
%
For two homogeneous slow-node settings, RWP-3m/s and Grid-3m/s, with $r=20$m (sparse connectivity), DRL-CL leads to significantly more forwards than Utility-based forwarding.
%with only 7\% and 4\% lower delay. 
The base model does not perform well for these two cases since the only sparse setting that it is trained on has mixed speed, instead of homogeneous speed. We therefore fine-tune the base model using 2000s of data from RWP-3m/s, $r=20$m. 
Specifically, we take
the base model as the initial model and then the 2000s of data to fine-tune the model in rounds (each round is set to 
50s).
%The results of the fine-tuned model are listed in bold for these two cases in Table~\ref{tb:testing-large-res}. 
We see that the fine-tuned model leads to similar delay as the base model, while significantly fewer forwards 
%than the base model 
(see bold results for these two cases in Table~\ref{tb:testing-large-res}). Using this fine-tuned model for these two mobility settings with $r=50$m also reduces the number of forwards, but it is less effective in highly connected scenarios (when $r=80$m).

To summarize, our approach is generalizable to the 36 scenarios Table~\ref{tb:testing-large-res}: it achieves 
lower 
%average 
delay than Utility-based forwarding, with reduction up to 
$78\%$ (specifically, 194s vs. 873s in RWP-3m/s, $r=80$m), except for one case where both schemes have low delay. 
%(leading to 28s more delay), with 
In addition, with fine-tuning using only 2000s of data in one case, it leads to comparable or slightly more forwards than Utility-based forwarding.

\begin{table*}[h]
    \centering
    \caption{
    Large-scale (100 nodes in 1000 m $\times$ 1000 m area) testing results. Each cell lists results in the order of Oracle, DRL-CL, Utility-based forwarding. In two rows,
    %(RWP-3m/s, Grid-3m/s), 
    results in bold are from the fine-tuned model using %2000s of data from 
    RWP-3m/s, $r=20$m.
    % Large-scale (100 nodes moving in 1000 m $\times$ 1000 m area) testing results. Each cell lists the results in the order of Oracle, DRL-CL, and Utility.
    }
    \begin{tabular}{l||c|c|c||c|c|c} % Changed from 'l|cccc' to 'l|ccc|ccc'
    \hline
    \multirow{2}{*}[-1.5pt]{\textbf{Setting}}  & \multicolumn{3}{c||}{\textbf{Delay (s)}} & \multicolumn{3}{c}{\textbf{\# of forwards}} \\ % Adjusted multicolumn spans and added labels
    \cline{2-4} \cline{5-7} % Adjusted \cline for the two groups
    & $r=20\,\text{m}$ & $r=50\,\text{m}$ & $r=80\,\text{m}$ & $r=20\,\text{m}$ & $r=50\,\text{m}$ & $r=80\,\text{m}$ \\ % Duplicated the r values
    \hline
    1-group & 113, 690, 1522 & 18, 63, 175 & 5, 61, 33 & 4, 67, 22 & 5, 22, 8 & 3, 40, 4 \\ % Added placeholder data
    \hline
    2-group & 124, 942, 1594 & 24, 67, 254 & 6, 38, 52 & 4, 56, 21 & 6, 25, 9 & 4, 22, 4 \\ % Added placeholder data
    \hline
    RWP-3m/s & 342, 5138 {\bf (4853)}, 5542 & 167, 1324 {\bf (1113)} , 2261 & 51, 194, 873 & 3, 120 {\bf (10)}, 27 & 4, 59 {\bf (31)}, 26 & 8, 35, 17 \\ % Added placeholder data
    RWP-15m/s & 66, 1029, 1076 & 36, 332, 396 & 18, 142, 222 & 3, 5, 15 & 3, 4, 12 & 6, 5, 9 \\ % Added placeholder data
    RWP-mix1 (20,80) & 74, 1094, 1169 & 39, 352, 506 & 19, 148, 322 & 3, 5, 16 & 3, 4, 15 & 6, 5, 12 \\ % Added placeholder data
    RWP-mix2 (50,50) & 98, 1189, 1455 & 49, 380, 640 & 22, 159, 390 & 4, 5, 16 & 4, 4, 16 & 6, 6, 13 \\ % Added placeholder data
    RWP-mix3 (80,20) & 159, 1293, 2332 & 74, 419, 860 & 30, 160, 393 & 3, 4, 19 & 4, 6, 17 & 6, 9, 12 \\ % Added placeholder data
    \hline
    Grid-3m/s & 628, 7817 {\bf (7999)}, 8155 & 427, 3706 {\bf (4492)}, 6342 & 158, 561, 1924 & 4, 584 {\bf (7)}, 27 & 4, 688 {\bf (17)}, 41 & 7, 71, 27 \\ % Added placeholder data
    Grid-15m/s & 124, 1779, 1779 & 84, 984, 1073 & 40, 374, 564 & 4, 11, 17 & 4, 11, 19 & 6, 13, 15 \\ % Added placeholder data
    Grid-mix1 (20,80) & 135, 1884, 1975 & 92, 1060, 1278 & 44, 424, 625 & 4, 10, 18 & 4, 10, 21 & 6, 12, 16 \\ % Added placeholder data
    Grid-mix2 (50,50) & 171, 2038, 2494 & 117, 1177, 1575 & 53, 502, 714 & 4, 9, 19 & 4, 9, 22 & 6, 11, 16 \\ % Added placeholder data
    Grid-mix3 (80,20) & 267, 2203, 3581 & 182, 1306, 2081 & 79, 571, 887 & 4, 10, 20 & 4, 12, 22 & 6, 12, 17 \\ % Added placeholder data
    \hline
\end{tabular} \label{tb:testing-large-res}
\end{table*}

%wb, 5/23/2025, provide insights on what is preferred by the model
\tydubfigsingle{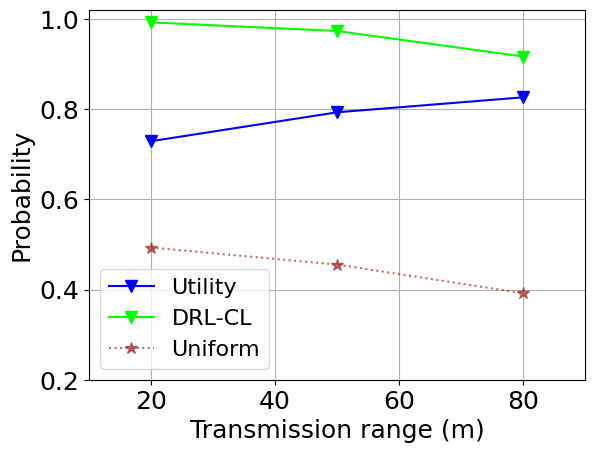}{RWP-mix3.}{figs/final_results/Probability-of-forwarding/test_2group_N100}{2-group.}{Forwarding preferences of different forwarding policies.
%(large networks). 
}{prob-forward-preference}

%%wb, 7/26/2025, comment out, do not remove
\iffalse 
\triplefig{figs/results/Probability of forwarding/test_RWP_mix_15_more_with_legend.png}{RWP-mix1}{figs/results/Probability of forwarding/test_RWP_mix_half.png}{RWP-mix2}{figs/results/Probability of forwarding/test_RWP_mix_3_more.png}{RWP-mix3}{Probability of forwarding to a fast node, RWP mixed settings. \bing{Three curves: Utility, DRL-CL, uniform.}}{prob-fast-forward-RWP}  

\triplefig{figs/results/Probability of forwarding/test_Manhattan_20x20_1000x1000_mix_15more_with_legend.png}{Grid-mix1}{figs/results/Probability of forwarding/test_Manhattan_20x20_1000x1000_mix}{Grid-mix2}{figs/results/Probability of forwarding/test_Manhattan_20x20_1000x1000_mix_3more}{Grid-mix3}{Probability of forwarding to a fast node, Grid mixed settings. \bing{Place holder.}}{prob-fast-forward-Grid} 

\smsinglefig{figs/results/Probability of forwarding/test_2group_N100_with_legend}{Probability of forwarding to a node in the destination group (2-group setting).\bing{Place holder.}}{2-group-pref}
\fi 

%\subsection{Learned policy} 
%\bing{7/18/2025} Insights on the behavior of the learned policy. Each plot with two curves: behavior of Utility and CL2. 

%We consider only those forwarding decisions that do not result in the packet being delivered, since there are more slow nodes, so there will be more traffic to be delivered to slow nodes which will skew statistics.

{\em Forwarding behavior.} 
%We next provide insights into the behavior of DRL-CL. 
In mixed-speed  scenarios with sparse connectivity, intuitively, it is generally better to forward packets to faster nodes, as they can cover greater distances more quickly, enabling faster delivery to the destination. We examine whether DRL-CL has this behavior.
%
%For mixed-speed scenarios, when the network is sparse, intuitively, forwarding packets to fast nodes is preferable, since such nodes can carry the packets to farther distances, and hence reach the destination faster.   We examine whether DRL-CL prefers forwarding to fast nodes. 
Specifically, for all decision makings %events 
when a node's neighbors contain both fast and slow nodes, we obtain the probability that DRL-CL chooses a fast node. For comparison, we also obtain the corresponding probability for Utility-based forwarding and when following a uniform random policy. 
%In addition, we obtain the probability that a fast node is chosen following a uniform random policy. 
We find that DRL-CL indeed prefers fast nodes. Fig.~\ref{fig:prob-forward-preference}a shows an example for RWP-mix3. We see DRL-CL has higher probability of choosing fast nodes than Utility-based strategy and uniform choice, and the probability decreases with the transmission range, while Utility-based strategy shows the opposite trend.
Similarly, for 2-group scenarios, we examine whether DRL-CL prefers forwarding to a neighbor that is in the same group as the destination. Fig.~\ref{fig:prob-forward-preference}b shows that DRL-CL has slightly higher probability of choosing a neighbor in the destination group than the uniform choice.

\triplefig{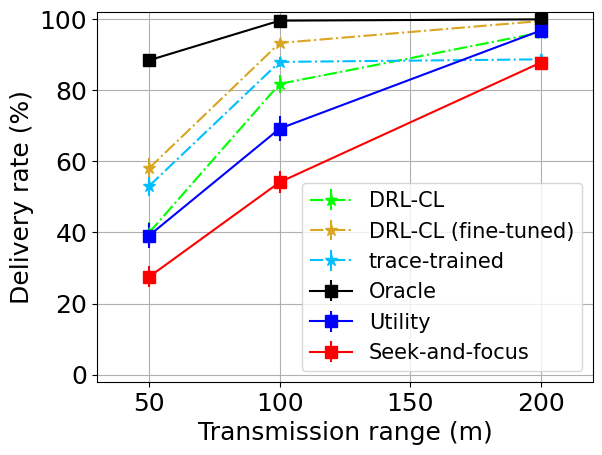}{Delivery rate.}{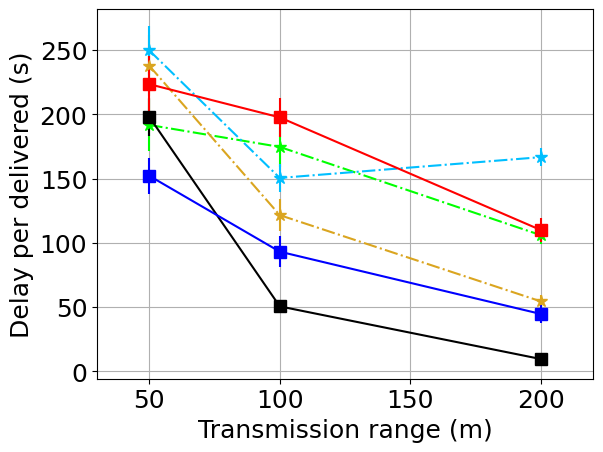}{Delay.}{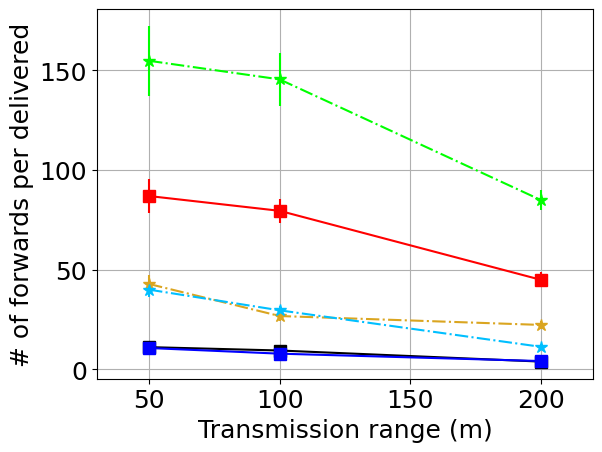}{\# of forwards.}{Results for a real-world scenario in Bologna (BO-1). }{Bologna-1-res}

\begin{table*}[h]
    \centering
    \caption{
    Results for real-world scenarios in Bologna (BO) and Bad Hersfelt (BA). Nodes move in 1500m$\times$1500m areas.
    %real-world scenarios (nodes moving in 1500m$\times$1500m areas). 
    %(nodes moving in 1500 m $\times$ 1500 m area) testing results. 
    Each cell lists the results in the order of Oracle, DRL-CL (fine-tuned), and Utility. The fine-tuning uses a single scenario, BO-1, $r=100\,\text{m}$. }
    \begin{tabular}{lc||c|c|c||c|c|c||c|c|c} % Changed from 'l|ccc|ccc' to 'l|cccc|cccc'
    \hline
    \multirow{2}{*}[-1.5pt]{\textbf{Setting}} & \multirow{2}{*}[-1.5pt]{\textbf{$N$}} &\multicolumn{3}{c||}{\textbf{Delivery rate (\%)}}& \multicolumn{3}{c||}{\textbf{Delay (s)}} & \multicolumn{3}{c}{\textbf{\# of forwards}} \\ % Adjusted multicolumn spans for 4 columns each
    \cline{3-6} \cline{7-11} % Adjusted \cline for the two groups with 4 columns
    & & $r=50\text{m}$ & $r=100\text{m}$ & $r=200\text{m}$ & $r=50\text{m}$ & $r=100\text{m}$ & $r=200\text{m}$ & $r=50\text{m}$ & $r=100\text{m}$ & $r=200\text{m}$ \\
    \hline
    BO-1 & 238 & 88, 58, 39 & 100, 93, 69 & 100, 99, 97 & 198, 238, 152 & 51, 122, 93 & 10, 54, 45 & 11, 43, 11 & 10, 27, 8 & 4, 22, 4 \\ % Updated placeholder data to integers
    BO-2 & 213 & 87, 51, 41 & 100, 90, 76 & 100, 99, 99 & 212, 160, 140 & 68, 156, 129 & 6, 69, 41 & 10, 30, 11 & 9, 26, 9 & 4, 27, 4 \\ % Updated placeholder data to integers
    BO-3 & 135 & 94, 66, 52 & 100, 97, 93 & 100, 100, 100 & 158, 212, 174 & 42, 70, 71 & 9, 41, 14 & 10, 39, 11 & 7, 14, 7 & 3, 17, 3 \\ % Updated placeholder data to integers
    \hline
    BA-1 & 53 & 60, 40, 39 & 86, 77, 67 & 100, 96, 91 & 177, 160, 153 & 114, 117, 100 & 47, 59, 48 & 4, 15, 5 & 4, 10, 4 & 3, 5, 3 \\ % Updated placeholder data to integers
    BA-2 & 49 & 59, 33, 35 & 74, 60, 54 & 100, 94, 90 & 228, 167, 184 & 155, 149, 148 & 72, 88, 78 & 3, 13, 3 & 4, 14, 4 & 3, 12, 3 \\ % Updated placeholder data to integers
    BA-3 & 44 & 59, 42, 34 & 73, 64, 59 & 97, 92, 89 & 198, 172, 164 & 111, 92, 87 & 49, 55, 54 & 3, 14, 4 & 4, 7, 5 & 3, 8, 3 \\ % Updated placeholder data to integers
    \hline
\end{tabular} \label{tb:real-world-res}
\end{table*}

\section{Handling Real-world Scenarios} \label{sec:real-world}

%Bologna: (0.0, 1496.25) (0.0, 1276.3); the other is very large, so clip 1500x1500 area

We now evaluate DRL-CL in real-world scenarios, where nodes (e.g., pedestrians and vehicles) navigate along urban road networks. For this purpose, we use Simulation of Urban MObility (SUMO)~\cite{Behrisch11:SUMO-overview}, which
%a %simulation 
%tool developed by transportation scientists. 
%This tool 
inputs realistic information (e.g., road networks, infrastructure, traffic lights) to simulate urban traffic based on transportation science. We consider road networks in two cities, one for 
%the city of 
Bologna, Italy~\cite{Bieker2014Traffic} and the other for 
%the city of 
Bad Hersfelt, Germany~\cite{Gotzler2022Assessment}. 

The road network for Bologna spans an area of roughly 1500m$\times$1500m. Since a node can move out of the network, we choose intervals of 1000 seconds and only consider the nodes that remain in the network throughout the interval. By varying the time intervals, we obtained three scenarios (BO-1, BO-2, BO-3), with the three highest number of nodes ($N=135$ to $N=238$ nodes).
%, listed in the first three rows in Table~\ref{tb:use-cases}. 
%\footnote{The last 300 s (of the total 1000 s duration) is the cool-down period.}. 
%
%We set the transmission range to 50, 100, or 200m, leading $3\times3=9$ scenarios, with average node degree from 5.4 to 35.4.   
The road network for Bad Hersfelt spans a much larger area. We select areas of size 1500m$\times$1500m and vary the time to select three scenarios (BA-1, BA-2, BA-3) with the most nodes ($N=44$ to $N=53$ nodes).
%; see the last three rows of Table~\ref{tb:use-cases}. 
The transmission range is set to 50 to 200m, leading to  $3\times3=9$ scenarios for each city, with average node degree 5.4 to 35.4 for Bologna, and from 1.8 to 10.2 for Bad Hersfelt. For each scenario, we generate packets among nodes 
using a similar process 
as in \S\ref{sec:choose-scenario-CL}.

{\em Fine-tuning base model.} We find that the base model does not perform well in the above scenarios. This is perhaps not surprising since it is trained using synthetic mobility models, which differs significantly from real-world mobility in road networks. We therefore fine-tune the base model using one trace of duration $1000$s for Bologna (BO-1), $r=100$m. Specifically, we take the base model as the initial model and fine-tune it using 1000s of packet trace from BO-1.
%gathered from 
%Bologna-1 to fine tune the model in rounds (each round is 50s). 

{\em Testing results.} We test this fine-tuned model in the 18 scenarios across the two cities. For each scenario, we generate 30 packet traces  and obtain the average results with 95\% confidence intervals. The duration of each trace is 1000s (the last 300s is the cool-down period). We next detail the results for BO-1, and then the results for all 
%the 18 
scenarios. 

Fig.~\ref{fig:Bologna-1-res} shows the result for BO-1. The delivery rate is below 100\% for all the schemes due to the short interval (1000s). DRL-CL (fine-tuned) outperforms DRL-CL (i.e., without fine-tuning), and leads to much higher delivery rate than Utility-based forwarding, which 
%and Seek-and-focus. In general, Utility-based forwarding 
further outperforms Seek-and-focus (we omit the results for Seek-and-focus in the following due to its generally worse performance). Since DRL-CL (fine-tuned) achieves higher delivery rates than Utility-based forwarding, it is reasonable that DRL-CL (fine-tuned) leads to higher delays and more forwards---in fact, sometimes Oracle scheme has higher delays than Utility-based forwarding as well. 
%only present the results with around 10 to 30 more forwards.
%
%For comparison, 
We further train a DRL model directly using the data that was used for fine-tuning the base model. The test results for this model are marked as `trace-trained' in Fig.~\ref{fig:Bologna-1-res}. We see that it does not generalize well: it leads to much lower delivery rate when $r=200$m (even lower than that when $r=100$m) than other schemes. This is perhaps not surprising since it is trained only using a small amount (1000s) of data.
%, which is often the case in realistic scenarios. 
In contrast, this small amount of data is sufficient for our fine-tuning since it %leverages 
builds on the knowledge embedded in the base model.

Table~\ref{tb:real-world-res} lists the results using the above fine-tuned model for all the 18 scenarios across the two cities. 
%We find that the above fine-tuned model using a single scenario generalizes well to all the   
It achieves up to 24\% (specifically, 93\% vs. 69\% for BO-1, $r=100$m) higher delivery rate than Utility-based forwarding. In the cases where these two schemes have similar delivery rate (BO-1, BO-2, BO-3, when $r=200$m), the fine-tuned model has slightly higher delay and more forwards. Overall,  our approach generalizes well to the 18 real-world scenarios.

%\input{related-work}
%\mysection{Conclusions and Future Work}
%\vspace{-0.1in}
\section{Conclusions}
\label{sec:conclusions}

In this paper, we have developed a framework for training generalizable DRL-based forwarding strategies for mobile wireless networks. We proposed three new classes of features for characterizing diverse networks and a CL-based approach for adapting forwarding to diverse networks. Extensive evaluation in a wide range of scenarios, including real-world road networks, demonstrates that our approach leads to generalizable forwarding strategies.

\iffalse 
We have designed and evaluated a DRL-based multi-copy forwarding strategy that can be practically constructed from a single-copy forwarding strategy. We have also proposed temporal features to better characterize sparse networks and give ways to estimate these features in a distributed way.  We evaluated our forwarding strategy in a variety of network scenarios, and showed that this strategy can perform significantly better than state-of-art multi-copy forwarding strategies by achieving consistently low delivery latency  and number of forwards across diverse network settings. We also showed the benefits of using temporal features in addition to spatial features to reduce unnecessary forwards.  Finally, our results demonstrate that very few copies are needed to significantly reduce delay, with the first copy giving the largest reduction in delay per packet delivered in the scenarios we consider.  In future work, we are interested in exploring sparse heterogeneous network scenarios, designing new features and spraying methodologies for these settings as well as formulating new learning problems. 
\fi 

%\bibliographystyle{ACM-Reference-Format}
%\bibliographystyle{abbrv}
\bibliographystyle{IEEEtran}
\bibliography{routing,bing-ref,RLReferences,xiaolan-ref}

\end{document}